\documentclass[aps,prb,twocolumn,superscriptaddress]{revtex4-2}
\usepackage{amsfonts}
\usepackage{graphicx}
\usepackage{epstopdf}
\usepackage{epsfig}
\usepackage{amsmath}
\usepackage[normalem]{ulem}
\usepackage{multirow}
\usepackage{makecell}
\usepackage{array}
\usepackage{booktabs}
\usepackage{mathtools}
\usepackage{bm}
\usepackage{color}
\usepackage{soul,xcolor}
\usepackage{array}
\usepackage{booktabs}
\usepackage{blindtext}

\usepackage{tabularx}
\newcolumntype{Y}{>{\centering\arraybackslash}X}

\begin{document}
\setstcolor{red}

\title{Irregular Bloch-Zener oscillations in $\alpha$-$\mathcal{T}_3$ lattices} 

\author{Li-Li Ye }
\affiliation{School of Electrical, Computer and Energy Engineering, Arizona State University, Tempe, Arizona 85287, USA}



\author{Ying-Cheng Lai} \email{Ying-Cheng.Lai@asu.edu}
\affiliation{School of Electrical, Computer and Energy Engineering, Arizona State University, Tempe, Arizona 85287, USA}
\affiliation{Department of Physics, Arizona State University, Tempe, Arizona 85287, USA}

\date{\today}

\begin{abstract}

	When a static electrical field is applied to a two-dimensional (2D) Dirac material, Landau-Zener transition (LZT) and Bloch-Zener oscillations can occur. Employing $\alpha$-$\mathcal{T}_3$ lattices as a paradigm for a broad class of 2D Dirac materials, we uncover two phenomena. First, due to the arbitrarily small energy gaps near a Dirac point that make it more likely for LZTs to occur than in other regions of the Brillouin zone, the distribution of differential LZT probability in the momentum space can form a complicated morphological pattern. Second, a change in the LZT morphology as induced by a mutual switching of the two distinct Dirac points can lead to irregular Bloch-Zener oscillations characterized by a non-smooth behavior in the time evolution of the electrical current density associated with the oscillation. These phenomena are due to mixed interference of quantum states in multiple bands modulated by the geometric and dynamic phases. We demonstrate that the adiabatic-impulse model describing Landau-Zener-St\"{u}ckelberg interferometry can be exploited to calculate the phases, due to the equivalence between the $\alpha$-$\mathcal{T}_3$ lattice subject to a constant electrical field and strongly periodically driven two- or three-level systems. The degree of irregularity of Bloch-Zener oscillations can be harnessed by selecting the morphology pattern, which is potentially experimentally realizable.  

\end{abstract}

\date{\today}

\maketitle

\section{Introduction} \label{sec:intro}

The Landau-Zener transition (LZT)~\cite{landau:1932,zener:1932} is a 
fundamental phenomenon in time-dependent quantum systems. The paradigmatic 
setting for LZT is a two-level system in which the two energy levels do not 
cross each other and vary adiabatically with time. When the energy gap between 
the two levels is sufficiently small, a non-adiabatic transition from one 
energy level to another can occur - leading to an LZT. The phenomenon of LZT 
is relevant to quantum information science and technology, because qubits are 
essentially two-level systems~\cite{Kok:2007,graham:2022,berke:2022}. In 
addition to quantum systems, LZTs can arise in other physical situations such 
as optical lattices~\cite{Niu:1996,Salger:2007,Liu:2021} and electromechanical 
systems~\cite{kervinen:2019,ivakhnenko:2018}. When a two-level system is 
periodically driven by an electric field, the transition probability will 
depend on the phase accumulated by the two energy bands between subsequent 
crossings, leading to the so-called Landau-Zener-St\"{u}ckelberg 
interferometry~\cite{shevchenko:2010}. In general, interference among the 
quantum states in different energy bands is determined by two phases: 
geometric and dynamic, which correspond to the adiabatic and Stokes phases,
respectively, in the adiabatic-impulse model~\cite{shevchenko:2010,Damski:2006}
underlying the Landau-Zener-St\"{u}ckelberg interferometry, where the sum of 
the adiabatic and Stokes phases gives the St\"{u}ckelberg phase - a concept 
originated from strongly periodically driven two-level systems. A 
generalization from the two-level setting is the three-level LZT 
model~\cite{carroll:1986} with an additional flat 
band~\cite{khomeriki:2016,parkavi:2021}. 

In solid state physics, Bloch oscillations~\cite{bloch:1928,zener:1934} are
a fundamental phenomenon closely related to LZT, which occur when a static 
electric field is applied to a periodic lattice, leading to a linear increase 
with time in the electron momentum and generating a time-dependent quantum 
system. The basic periodicity of the momentum space stipulates that the 
electron must execute oscillatory motions in the physical space at a frequency 
determined by the lattice constant and the electric field strength (typically 
in the terahertz regime). In fact, insofar as the electron moves in a periodic 
potential, Bloch oscillations can occur, rendering them a common quantum 
phenomenon beyond a solid-state lattice. In the past, the oscillations have 
been observed in diverse systems such as semiconductor 
superlattices~\cite{waschke:1993}, photonic 
structures~\cite{morandotti:1999,sapienza:2003,Trompeter:2006,ghulinyan:2005} 
and plasmonic waveguide arrays~\cite{block:2014}. The phenomenon provides
a viable way to convert a direct current to a high-frequency 
signal~\cite{schubert:2014,fahimniya:2021}. 

Bloch oscillations arise from the time evolution of the electron in a single 
energy band. When there are multiple energy bands, LZTs can occur at the
avoided crossing points between the bands. Driven by the static electric field,
a quantum state initialized in the lower energy band evolves with time. 
At certain time, the state will reach an avoided crossing point between
distinct energy bands and possibly experience an LZT. Thus, in systems with
multiple energy bands, a combination of LZTs and Bloch oscillations can occur,
leading to the so-called Bloch-Zener 
oscillations~\cite{dreisow:2009,lim:2012,parkavi:2021,zhang:2021}, which have
applications in, e.g., matter-wave beam splitters and Mach-Zender
interferometry~\cite{breid:2006,breid:2007}. In the past, Bloch-Zener 
oscillations were extensively studied in one-dimensional (1D) gapped periodic 
lattices~\cite{morsch:2001,breid:2006,breid:2007,venkatesh:2009} and were 
demonstrated to be sensitive to the size of the energy gap. In particular, if 
the gap is relatively large, LZTs are inhibited so the Bloch-Zener oscillations
are restricted to within a single energy band. For a smaller gap, interband 
LZTs can occur, which destroys the periodicity of the Bloch-Zener 
oscillations~\cite{morsch:2001,breid:2006,breid:2007,lim:2015,kling:2010}. 
In a 2D lattice with multiple energy bands, in the first Brillouin zone, 
different subregions can arise that can inhibit or excite LZTs, resulting in 
irregular Bloch-Zener oscillations~\cite{sun:2018,chang:2022}. For example, in 
pure graphene, when a constant electric field is applied in a lattice-rational 
direction, the oscillation amplitude can decay with time rapidly in a 
non-adiabatic fashion~\cite{rosenstein:2010}. When the electric field is 
in an irrational direction with respect to the lattice structure, Bloch 
oscillations of complicated patterns can arise~\cite{kolovsky:2013}.

In this paper, we study the effects of a constant and uniform electric field 
on a broad class of 2D Dirac materials, the $\alpha$-$\mathcal{T}_3$ 
lattice~\cite{RMFPM:2014,ICN:2015}, which has an additional atom at the center 
of each unit cell of the honeycomb graphene lattice. The interaction between
the central atom and any of its nearest neighbors is characterized by the 
parameter $0 \le \alpha \le 1$ - effectively the strength relative to that 
between two neighboring atoms at the vertices of the graphene cell. For 
$\alpha=0$, the lattice reduces to that of graphene with quasiparticles being
pseudospin-1/2 Dirac fermions. As $\alpha$ increases from zero, a flat band 
through the conic interaction of the two Dirac cones 
emerges~\cite{RMFPM:2014,ICN:2015}. The maximal value $\alpha=1$ gives a 
pseudospin-1 lattice where, because of the extra atom, the low energy 
excitations need to be described by the pseudospin-1 Dirac-Weyl equation with 
a three-component spinor~\cite{BUGH:2009}. A distinct feature of the entire 
spectrum of $\alpha$-$\mathcal{T}_3$ lattices is the existence of two distinct 
valleys centered about the two non-equivalent Dirac points of the backbone 
hexagonal lattice, denoted as $+\mathbf{K}$ and $-\mathbf{K}$. 

\begin{figure} [ht!]
\centering
\includegraphics[width=\linewidth]{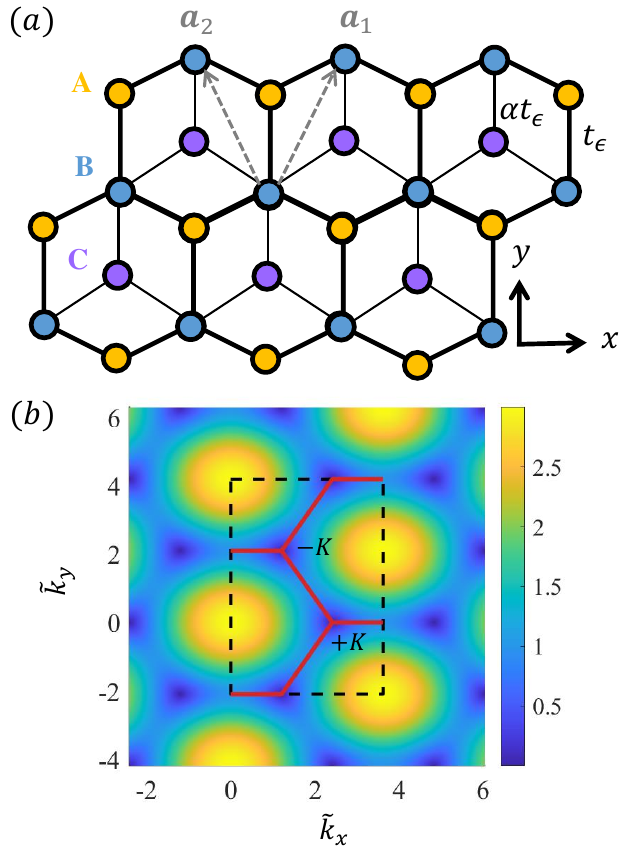}
\caption{Illustration of the $\alpha$-$\mathcal{T}_3$ lattice and its positive 
energy band structure. (a) The lattice structure~\cite{raoux:2014,kittel:2018} 
as defined by the three base atoms (A, B and C) in physical space spanned by 
the two primitive translation vectors $\pmb{a}_{1}$ and $\pmb{a}_{2}$. The 
nearest-neighbor hopping energy between A and B sites is $t_{\epsilon}$ and 
that between B and C sites is $\alpha t_{\epsilon}$, where $0\le \alpha \le 1$ 
characterizes the coupling strength. (b) Zero-field energy-band 
structure~\cite{raoux:2014} of the positive dispersion band as a function of 
the wave vector $\pmb{k}$ for an arbitrary value of $\alpha$ in the hexagonal 
Brillouin zone. The zero energy points correspond to two classes of 
non-equivalent contact points at the corners $\xi \mathbf{K}$ with the valley 
index $\xi=\pm 1$. A reference rectangular region (dashed line) for numerical 
integration is indicated, with the boundaries of the hexagonal Brillouin 
zone specified by the red solid lines.}
\label{fig:alpha_T3_lattice}
\end{figure}

Depending on the direction and magnitude of the electric field, electrons 
initiated from distinct valleys can exhibit characteristically different LZTs.
As the dynamic phases associated with different valleys cancel each other
exactly, the distinct LZTs are due to the different adiabatic phases of the 
quantum states in the energy bands between consecutive crossings. This can 
be understood by considering the reciprocal periodic momentum space with the 
hexagonal Brillouin zone, as shown in Fig.~\ref{fig:alpha_T3_lattice}. Now
apply an electric field in the $x$ direction. The $x$ component of the momentum
will then increase linearly with time under the premise of the same energy 
value. As a result, the Dirac points $+\mathbf{K}$ and $-\mathbf{K}$ will shift
towards the right. At an original Dirac point (+$\mathbf{K}$ or $-\mathbf{K}$),
the energy will increase from zero, reach a maximum, and then decrease to zero 
when the next Dirac point arrives, generating a time-periodic behavior. Because
of the hexagonal structure of the momentum space, the energy variations 
associated with $+\mathbf{K}$ and $-\mathbf{K}$ are distinct, as indicated in
Fig.~\ref{fig:alpha_T3_lattice}(b). The LZT probability depends on the 
accumulated phase between subsequent crossings, where the adiabatic phase is 
the integral of the energy variation over time. In the specific setting of 
Figs.~\ref{fig:alpha_T3_lattice}(a) and \ref{fig:alpha_T3_lattice}(b), the 
integral associated with the Dirac point $+\mathbf{K}$ will have a much larger 
value than that associated with the other Dirac point $-\mathbf{K}$. As a 
result, if an electron initiates with a momentum value near $+\mathbf{K}$, 
the adiabatic phase will be nearly constant for a large range of energy gaps
determined by the momentum deviation from the trajectory of Dirac points 
$\pm K$ in the $p_y$ direction. However, if an electron starts with a momentum 
value near $-\mathbf{K}$, the adiabatic phase will depend sensitively on the 
energy gap. Based on this property, it is possible to generate specific 
destructive or constructive interference for a large range of momentum 
deviation from the trace of the Dirac point $+K$, whereas there is mixed
interference associated with all possible phases for electrons with initial 
momentum near $-K$. 

Our first finding is the emergence of complicated LZT morphological patterns 
in the vicinity of distinct Dirac points, which is associated with mixed 
quantum interference among the quantum states in multiple bands. Say we apply 
an electric field in the $x$ direction, initialize electrons in the lower 
Dirac cone (the lower band), and calculate the differential LZT probability, 
defined as the difference between the probability that an electron is in the 
upper band and that in the lower band. Different momentum values about 
a Dirac point and the magnitudes of the electric field give distinct 
interference phases. As a result, in the momentum plane, the differential LZT 
probability displays different values, giving rise to some morphological 
pattern that can be complex. During its temporal evolution, the pattern can be 
maintained for some time but it can change from time to time due to the 
switching of two two distinct Dirac points $\pm\mathbf{K}$ in the Brillouin 
zone as caused by the external electric field and the periodic structure of 
the momentum space. Specifically, during one period of the Bloch-Zener 
oscillation, the two valleys go through a complete cycle in the sense that 
they are switched and then returned to their respective original positions. 

The second finding is that changes in the LZT morphology can lead to 
irregularities in Bloch-Zener oscillations in $\alpha$-$\mathcal{T}_3$ lattice.
To explain this, we recall the two typical cases where periodic Bloch 
oscillations are generated. One case is a single-band material, such as a 
normal conductor, where the Bloch oscillations are characterized by a perfect 
temporally periodic behavior in the electrical current density. Another case 
is where an LZT causes all electrons initialized in one band to transition 
completely to another band, i.e., the transition probability is one - the 
so-called ideal LZT. In this case, the resulting Bloch-Zener oscillations 
behave as if the electrons were in a single band. In the three-band 
$\alpha$-$\mathcal{T}_3$ lattice, LZTs are typically not ideal. The mutual 
switchings of the two Dirac points $\pm\mathbf{K}$ in the Brillouin zone 
changes the LZT morphology, which can produce an abrupt, nonsmooth change 
in the current density, thereby leading to aperiodic, irregular Bloch-Zener 
oscillations. More specifically, the coexistence of a variety of LZT 
possibilities in the momentum space generates complex, mixed quantum 
interference between the states in the upper, lower, and flat bands, 
disrupting the originally periodic Bloch-Zener oscillation rhythm before the 
Dirac point switch. While aperiodic Bloch
oscillations~\cite{morsch:2001,breid:2006,breid:2007,lim:2015,kling:2010}
and irregular Bloch-Zener oscillations~\cite{sun:2018,chang:2022} have been
noted before, to our knowledge, the physical mechanisms underlying these 
irregular behaviors were not clear. Especially, it has not been reported 
previously that LZTs can form a complicated morphology in 
$\alpha$-$\mathcal{T}_3$ lattice and a change in the morphology can lead to 
irregular Bloch-Zener oscillations.

In Sec.~\ref{sec:model}, we describe the $\alpha$-$\mathcal{T}_3$ lattice 
model and derive the current density associated with the Bloch-Zener 
oscillations from the adiabatic basis in the hexagonal Brillouin zone. In 
Sec.~\ref{sec:Irregular_BZO}, we present a general treatment of LZTs in the 
$\alpha$-$\mathcal{T}_3$ lattice, display the LZTs morphology in the long 
time, and analyze the relationship between morphology and irregular Bloch-Zener
oscillations. In particular, in Sec.~\ref{subsec:LZT_probability}, we linearize
the Hamiltonian about the Dirac points to obtain the effective Landau-Zener 
Hamiltonian for the two limiting cases: $\alpha=0$ and $\alpha=1$. For 
$0 < \alpha < 1$, we numerically demonstrate the occurrence of LZT. In 
Sec.~\ref{subsec:morphology_irregularBZO}, we elucidate the interplay between 
LZT morphological changes and irregular Bloch-Zener oscillations. In 
Sec.~\ref{sec:LZSI}, we focus on the Landau-Zener-St\"{u}ckelberg 
interferometry in $\alpha$-$\mathcal{T}_3$ lattice, where in 
Sec.~\ref{subsec:two_three_LZSI}, we establish the equivalence of the 
$\alpha$-$\mathcal{T}_3$ lattice to two- or three-level time-dependent systems 
and provide an understanding of the LZT based on the St\"{u}ckelberg phase. 
In Sec.~\ref{subsec:design_morphology}, we address the problem of harnessing
irregular Bloch-Zener oscillations through selection of the LZT morphology 
and discuss the experimental feasibility of this scheme. A discussion 
is offered in Sec.~\ref{sec:discussion}. Our main code is uploaded to 
GitHub: https://github.com/liliyequantum/Irregular-Bloch-Zener-oscillations-in-two-dimensional-flat-band-Dirac-materials.

\section{Basics of $\alpha$-$\mathcal{T}_3$ lattice, Landau-Zener transition, and Bloch-Zener oscillations} \label{sec:model}

The $\alpha$-$\mathcal{T}_3$ lattice interpolates between the graphene 
honeycomb lattice ($\alpha=0$) and the dice lattice ($\alpha=1$) with the 
parametrization $\tan\varphi=\alpha\in[0,1]$ with the duality~\cite{raoux:2014}
$\alpha\rightarrow 1/\alpha$. The tight-binding Hamiltonian is given by
\begin{equation}
    H=\left[\begin{array}{ccc}
0 & f_{\pmb{k}}\cos\varphi & 0\\
f_{\pmb{k}}^{*}\cos\varphi & 0 & f_{\pmb{k}}\sin\varphi\\
0 & f_{\pmb{k}}^{*}\sin\varphi & 0
\end{array}\right],
\end{equation}
where 
\begin{align} 
f_{\pmb{k}}=-t_{\epsilon}\left(1+e^{-i\pmb{k}\ensuremath{\cdot}\pmb{a}_{1}}+e^{-i\pmb{k}\cdot\pmb{a}_{2}}\right), 
\end{align}
$\pmb{k}=(k_{x},k_{y})$, and $t_{\epsilon}$ is the nearest-neighbor hopping energy 
between A and B sites, as shown in Fig.~\ref{fig:alpha_T3_lattice}(a). The 
primitive translation vectors are $\pmb{a}_{1}=a\left(\sqrt{3}/2,3/2\right)$ 
and $\pmb{a}_{2}=a\left(-\sqrt{3}/2,3/2\right)$ with $a$ being the lattice 
constant. The corresponding primitive translation vectors in the hexagonal
Brillouin zone of the reciprocal lattice are 
$\pmb{b}_{1}=\left(\sqrt{3}/3,1/3\right)2\pi/a$ and 
$\pmb{b}_{2}=\left(-\sqrt{3}/3,1/3\right)2\pi/a$. 
The eigenenergy spectrum of the $\alpha$-$\mathcal{T}_3$ lattice is independent
of $\alpha$, which consists of two conic dispersive bands 
$\epsilon_{\lambda}=\lambda\left|f_{\pmb{k}}\right|$ distinguished by the band index 
$\lambda=\pm$ and a zero energy flat band $\epsilon_{0}=0$, as shown in 
Fig.~\ref{fig:alpha_T3_lattice}(b). The eigenstates of the 
$\alpha$-$\mathcal{T}_3$ lattice in the whole hexagonal Brillouin zone can be 
obtained through the following effective Hamiltonian about the Dirac points:
\begin{align}
    |\psi_{0}\rangle=\left[\begin{array}{c}
\sin\varphi\;e^{i\theta_{\pmb{k}}}\\
0\\
-\cos\varphi\;e^{-i\theta_{\pmb{k}}}
\end{array}\right],
\;|\psi_{\lambda}\rangle=\frac{1}{\sqrt{2}}\left[\begin{array}{c}
\cos\varphi\;e^{i\theta_{\pmb{k}}}\\
\lambda\\
\sin\varphi\;e^{-i\theta_{\pmb{k}}}
\end{array}\right],
\end{align}
where $\theta_{\pmb{k}}$ is the angle of $f_{\pmb{k}}$ associated with the 
specific momentum. 

We describe the Hamiltonian underlying Bloch-Zener oscillations.
Apply a uniform and constant electric field to the $\alpha$-$\mathcal{T}_3$
lattice in the $+x$ direction, which is switched on at $t=0$. With the 
time-dependent vector potential~\cite{dora:2010,wang:2017} 
$\pmb{A}\left(t\right)=\left[A\left(t\right),0,0\right]$, where 
$A\left(t\right)=Et\Theta\left(t\right)/\hbar$, the Hamiltonian becomes
\begin{align}
    H\left(t\right)=\left[\begin{array}{ccc}
0 & f_{\pmb{k}}\left(t\right)\cos\varphi & 0\\
f_{\pmb{k}}^{*}\left(t\right)\cos\varphi & 0 & {\color{blue}f_{\pmb{k}}\left(t\right)\sin\varphi}\\
0 & f_{\pmb{k}}^{*}\left(t\right)\sin\varphi & 0
\end{array}\right],
\end{align}
where $f_{\pmb{k}}(t)$ is given by
\begin{align}
    f_{\pmb{k}}(t)=-t_{\epsilon}\left(1+2e^{-i\frac{3}{2}\widetilde{k}_{y}}\cos\left(\frac{\text{\ensuremath{\sqrt{3}}}}{2}\widetilde{k}_{x}\left(t\right)\right)\right),
\end{align}
with $k_{x}\left(t\right)\equiv k_{x}-eEt/\hbar$. For convenience, we define 
the dimensionless quantities 
\begin{align}
\widetilde{k}_{x}\left(t\right) &\equiv k_{x}\left(t\right)a, \\
\widetilde{k}_{y} &\equiv k_{y}a, 
\end{align}
so that $f_k(t)$ has the same dimension as the hopping energy $t_{\epsilon}$.

In the presence of the electric field, the eigenenergy spectrum of the 
positive dispersion band in the whole hexagonal Brillouin zone is determined by 
\begin{align} \nonumber
\epsilon_{\pmb{k}}(t) &=\left|f_{\pmb{k}}(t)\right| \\ \label{eq:energy}
	&=t_{\epsilon}\sqrt{1+4\cos X_{\pmb{k}}(t)\left(\cos Y_{\pmb{k}}+\cos X_{\pmb{k}}\left(t\right)\right)}
\end{align}
where $X_{\pmb{k}}\left(t\right)=\widetilde{k}_{x}\left(t\right)\sqrt{3}/2$ and
$Y_{\pmb{k}}=\widetilde{k}_{y}3/2$. Because the function $f_{\pmb{k}}(t)$ 
includes $\cos(L(t))$, where $L(t)$ is a linear function of time, the 
Hamiltonian becomes time-periodic. Consider the special case of $\alpha=0$ 
(graphene). Expanding the Hamiltonian around the Dirac points yields the 
standard Landau-Zener Hamiltonian~\cite{tayebirad:2010}:
\begin{align} \label{eq:LZ_Hamiltonian}
    \widetilde{H}\approx\frac{3\delta\widetilde{k}_{y}}{2}\sigma_{x}\mp\frac{3\widetilde{E}\widetilde{t}'}{2}\sigma_{z}.
\end{align}
That is, when the electrons are near the Dirac points, LZTs between distinct
energy bands can arise.

The quantum dynamics are governed by
\begin{align}
    i\hbar\partial_{t}\psi_{\pmb{k}}\left(t\right)=H\left(t\right)\psi_{\pmb{k}}\left(t\right).
\end{align}
On the adiabatic basis, the evolution of a quantum state is under an 
infinitesimal electric field~\cite{dora:2010,vitanov:1996}:
\begin{align}
    U^{\dagger}\left(t\right)H\left(t\right)U\left(t\right)=S_{z}\epsilon_{\pmb{k}}\left(t\right),
\end{align}
where $S_z$ is the $z$-component of the vector of spin-1 matrices. The 
transformed quantum dynamics are governed by
\begin{align}\label{eq:adiabatic_evolution}
    i\hbar\partial_{t}\Phi_{\pmb{k}}\left(t\right)=\biggl[S_{z}\epsilon_{\pmb{k}}\left(t\right)- \widetilde{S}_{x}\frac{at_{\epsilon}^{2}eE}{\epsilon_{\pmb{k}}^{2}\left(t\right)}C_{0}\left(t\right)\biggr]\Phi_{\pmb{k}}\left(t\right),
\end{align}
through the time-dependent unitary transformation $U(t)$ given by
\begin{align}
   \left[\begin{array}{ccc}
\frac{1}{\sqrt{2}}\cos\varphi\;e^{i\theta_{\pmb{k}}(t)} & \sin\varphi\;e^{i\theta_{\pmb{k}}(t)} & \frac{1}{\sqrt{2}}\cos\varphi\;e^{i\theta_{\pmb{k}}(t)}\\
\frac{1}{\sqrt{2}} & 0 & -\frac{1}{\sqrt{2}}\\
\frac{1}{\sqrt{2}}\sin\varphi\;e^{-i\theta_{\pmb{k}}(t)} & -\cos\varphi\;e^{-i\theta_{\pmb{k}}(t)} & \frac{1}{\sqrt{2}}\sin\varphi\;e^{-i\theta_{\pmb{k}}(t)}
\end{array}\right],
\end{align}
where $\Phi_{\pmb{k}}(t)=U^\dagger(t)\psi_{\pmb{k}}(t)$ and the term 
incorporating $C_0(t)$ contributes to the time dependence of $U(t)$ through
$-i\hbar U^{\dagger}\partial_{t}U$. Specifically, we have
\begin{align} 
C_{0}\left(t\right) &=\sqrt{3}\sin Y_{\pmb{k}}\sin X_{\pmb{k}}\left(t\right) \\
\widetilde{S}_{x} &=S_{x}\sin2\varphi-S_{L}\cos2\varphi, 
\end{align}
where 
\begin{align}
    S_{x}=\frac{1}{\sqrt{2}}\left[\begin{array}{ccc}
0 & 1 & 0\\
1 & 0 & 1\\
0 & 1 & 0
\end{array}\right],\;S_{L}&\equiv \left[\begin{array}{ccc}
-\frac{1}{2} & 0 & -\frac{1}{2}\\
0 & 1 & 0\\
-\frac{1}{2} & 0 & -\frac{1}{2}
\end{array}\right].
\end{align}
For $\alpha=1$, $\widetilde{S}_x$ reduces to $S_x$, the $x$ component of 
spin-1 matrix and Eq.~(\ref{eq:adiabatic_evolution}) becomes the quantum 
evolution equation for a dice lattice. This form of 
Eq.~(\ref{eq:adiabatic_evolution}) is consistent with that reported in 
a previous work~\cite{wang:2017} except for a periodic factor in $C_0(t)$ 
due to the intrinsic lattice structure. For $\alpha=0$, $\widetilde{S}_x$ 
become $-S_L$ and Eq.~(\ref{eq:adiabatic_evolution}) describes the graphene 
lattice, which is consistent with a previous work~\cite{dora:2010} except 
for the identity term $I\exp(-i\theta(t)/2)\partial_{t}\exp(i\theta(t)/2)/2$ 
due to the different form of the basis. Overall, $\widetilde{S}_x$ reflects 
the different coupling strength among the three bands and the periodic term 
in $C_0(t)$ originates from the intrinsic property of the 
$\alpha$-$\mathcal{T}_3$ lattice.

We set the initial state as one corresponding fully occupied lower band: 
\begin{align}
    \Phi_{\pmb{k}}\left(t=0\right)=\left[0,0,1\right]^{T}.
\end{align}
The average current density associated with the momentum 
$\langle J_{x}\rangle_{\pmb{k}}\left(t\right)$ in the hexagonal Brillouin zone 
is given by
\begin{align}
  \langle J_{x}\rangle_{\pmb{k}}\left(t\right)\equiv\Phi_{\pmb{k}}^{\dagger}\left(t\right)J_{x,\;\pmb{k}}\left(t\right)\Phi_{\pmb{k}}\left(t\right)
\end{align}
where the current density matrix with the definite momentum is 
\begin{align} 
J_{x,\;\pmb{k}}\left(t\right)=-eU^{\dagger}\left(t\right)\partial_{k_{x}\left(t\right)}H(t)U\left(t\right)
\end{align}
and $\Phi_{\pmb{k}}\left(t\right)$ in the adiabatic basis can be written as
\begin{align}
  \Phi_{\pmb{k}}\left(t\right)=\left[\begin{array}{ccc}
\alpha_{\pmb{k}}\left(t\right), & \gamma_{\pmb{k}}\left(t\right), & \beta_{\pmb{k}}\left(t\right)\end{array}\right]^{T}.
\end{align}
Due to the periodic structure of the energy band (referred to as the
Bloch band), the average current density will exhibit Bloch oscillations.

The average current density $\langle J_{x}\rangle_{\pmb{k}}\left(t\right)$ can 
be decomposed into two components, the intraband and interband currents, 
respectively~\cite{dora:2010}: 
\begin{align}\label{eq:J}
  \langle J_{x}\rangle_{\pmb{k}}\left(t\right)=\langle J_{x}\rangle_{\pmb{k}}^{intra}\left(t\right)+\langle J_{x}\rangle_{\pmb{k}}^{inter}\left(t\right),
\end{align}
which can be written as
\begin{align}\label{eq:J_intra_domain}
 &\langle J_{x}\rangle_{\pmb{k}}^{intra}\left(t\right)=J_{x,\;\pmb{k}}^{11}\left(t\right)\left(\left|\alpha_{\pmb{k}}\left(t\right)\right|^{2}-\left|\beta_{\pmb{k}}\left(t\right)\right|^{2}\right),\\
  &\langle J_{x}\rangle_{\pmb{k}}^{inter}\left(t\right)=2\Re\left[J_{x,\;\pmb{k}}^{13}\left(t\right)\alpha_{\pmb{k}}^{*}\left(t\right)\beta_{\pmb{k}}\left(t\right)\right]\nonumber\\
  &+2\Re\left[J_{x,\;\pmb{k}}^{12}\left(t\right)\alpha_{\pmb{k}}^{*}\left(t\right)\gamma_{\pmb{k}}\left(t\right)+J_{x,\;\pmb{k}}^{23}\left(t\right)\gamma_{\pmb{k}}^{*}\left(t\right)\beta_{\pmb{k}}\left(t\right)\right].
\end{align}
The matrix $J_{x,\pmb{k}}(t)$ provides insights into the current density 
$\langle J_x \rangle_{\pmb{k}}(t)$. In particular, the intraband component 
consists of both electrons and holes, corresponding to
\begin{align} 
J_{x,\;\pmb{k}}^{11}\left(t\right) &\equiv J_{x,\;\pmb{k}}^{0}\left(t\right)\cos\Theta_{\pmb{k}}\left(t\right) \\ 
J_{x,\;\pmb{k}}^{33}\left(t\right) &=-J_{x,\;\pmb{k}}^{11}\left(t\right), 
\end{align}
respectively, where the minus sign comes from the opposite sign of the 
equivalent charge in the electron-hole pair. The zero group velocity of the 
flat band results in zero intraband contribution. The interband contribution 
arises from the interference between the transition from the lower to the flat 
band or the upper band and that from the flat to the upper band, corresponding 
to $J_{x,\;\pmb{k}}^{23}\left(t\right),\;J_{x,\;\pmb{k}}^{13}\left(t\right)\;$ 
and $J_{x,\;\pmb{k}}^{12}\left(t\right)$, respectively, which are given by
\begin{align}
  J_{x,\;\pmb{k}}^{13}\left(t\right)&\equiv i J_{x,\;\pmb{k}}^{0}\left(t\right)\cos2\varphi\;\sin\Theta_{\pmb{k}}\left(t\right),\\
  J_{x,\;\pmb{k}}^{12}\left(t\right)=J_{x,\;\pmb{k}}^{23}\left(t\right)&\equiv i J_{x,\;\pmb{k}}^{0}\left(t\right)/\sqrt{2}\sin2\varphi\sin\Theta_{\pmb{k}}\left(t\right),
\end{align}
where 
\begin{align}
\Theta_{\pmb{k}}\left(t\right)\equiv\theta_{\pmb{k}}\left(t\right)+Y_{\pmb{k}}
\end{align}
and $J_{x,\;\pmb{k}}^{0}(t)$ is the common factor with the dimension of the 
current density: 
\begin{align}
J_{x,\;\pmb{k}}^{0}\left(t\right)=-\sqrt{3}eat_{\epsilon}\sin X_{\pmb{k}}\left(t\right).
\end{align} 
To facilitate numerical calculations, we define the dimensionless quantities:
\begin{align}
    \widetilde{t} &=t/t_0, \\
    \widetilde{E} &=E/E_0,\\
    \widetilde{\epsilon}_{k}(t) &=\epsilon_k(t)/\epsilon_0, \\
    \widetilde{J}_{x,\;\pmb{k}}(t) &=J_{x,\;\pmb{k}}(t)/J_0.
\end{align}
where $t_0 \equiv \hbar/t_{\epsilon}$, $E_0 \equiv t_{\epsilon}/(ea)$, $\epsilon_0 = t_{\epsilon}$ and 
$J_0 = e a t_{\epsilon}$. We use the fourth-order Runge–Kutta method to calculate the 
adiabatic evolution of the particle. The discrete step sizes in time and 
momentum are chosen according to Eq.~\eqref{eq:adiabatic_evolution} with
the error tolerance $10^{-2}$ and normalized wavefunction error within 
$10^{-4}$. Figure~\ref{fig:alpha_T3_lattice}(b) indicates that the 
$\pm\mathbf{K}$ correspond to zero energy, which results in numerical 
divergence of Eq.~(\ref{eq:adiabatic_evolution}), so we set a minimum energy 
cut-off to be $\widetilde{\epsilon}_{k}\geq 10^{-5}$.

The dimensionless current density $\widetilde{J}$ is the result of integrating 
$\langle J_x \rangle_{\pmb{k}} (t)/J_{0}$ over the rectangular reference area 
in the hexagonal Brillouin zone, as shown in Fig.~\ref{fig:alpha_T3_lattice}(b),
which contains three Dirac points and is $3/2$ times larger than the first 
Brillouin zone. In addition, we divide the current density $\widetilde{J}$ by 
the electric field $\widetilde{E}$ and a constant $3\pi^2/4$, which normalizes 
the current density in the weak field regime~\cite{rosenstein:2010} for 
$\alpha=0$.

\section{Morphological changes in LZTs and irregular Bloch-Zener oscillations} \label{sec:Irregular_BZO}

\subsection{Landau-Zener transition} \label{subsec:LZT_probability}

To gain insights, we first consider the special case $\alpha=0$ (graphene), 
where the Hamiltonian of the $\alpha$-$\mathcal{T}_3$ lattice linearized about 
the Dirac points $\pm \mathbf{K}$ corresponds to that of a standard two-level 
system. Using the unitary transformation 
\begin{align}
    U=\exp\left(-i\pi/4\sigma_{y}\right)\exp\left(-i\pi/4\sigma_{z}\right),
\end{align}
we can write the linearized Hamiltonian as (Appendix~\ref{app:equivalence})
\begin{align} \label{eq:linearized_H_graphene}
    U^{\dagger}\widetilde{H}U\approx\frac{3\delta\widetilde{k}_y}{2}\sigma_{x}\mp\frac{3\widetilde{E}\widetilde{t}'}{2}\sigma_{z}
\end{align}
where $\delta\widetilde{k}_y$ is an infinitesimal deviation from a Dirac point 
and $\widetilde{t}^{'}=0$ denotes the starting time from the Dirac point. 
Recall the standard Landau-Zener Hamiltonian for a two-level 
system~\cite{tayebirad:2010}:
\begin{align} \label{eq:standar_LZT}
H_{LZ}=(g/2)\sigma_{x}+(st)\sigma_{z},
\end{align}
with the two underlying adiabatic energy levels 
\begin{align} \label{eq:H_LZ_energy}
\widetilde{\epsilon}_{\pm}=\pm\frac{1}{2}\sqrt{\left(2st\right)^{2}+g^{2}},
\end{align}
where $g$ is the energy gap and $s$ is the slope of the two-level band 
about the LZT point. The linearized Hamiltonian for graphene in 
Eq.~(\ref{eq:linearized_H_graphene}) can thus be cast in the standard two-level
LZT Hamiltonian with the following parameter correspondences: 
\begin{align} 
\alpha=0, \ g=3\delta\widetilde{k}_{y}, \ {\rm and} \ s=3\widetilde{E}/2. 
\end{align} 
For a finite $\Delta\widetilde{k}_{y}$ in $\alpha$-$\mathcal{T}_3$ lattice, 
the exact energy gap is 
\begin{align}\label{eq:gap_delta_k_y}
    g = 2\sqrt{1-\cos^2\left(\frac{3}{2}\Delta \widetilde{k}_y\right)}.
\end{align}
The gap size increases monotonically with the momentum deviation 
$\Delta\widetilde{k}_y$ from a Dirac point. [For $\alpha \ne 0$ (a flat band),
the gap between the lower band and the flat one is half of the gap between the 
lower and the upper bands.] For $\Delta\widetilde{k}_{y}=\pi/3$, the gap size 
$g$ reaches the maximum value of two in the energy unit $\epsilon_0$ in the 
hexagonal Brillouin zone.

In a two-level quantum system, for electrons initialized in the lower band, 
after the electric field is turned on, the first LZT to the upper band occurs 
with the probability~\cite{tayebirad:2010}
\begin{align}\label{eq:formula_first_LZT_up}
   P_{LZ}\equiv|\alpha_{\pmb{k}}|^2=\exp\left(-\pi r\right),
\end{align}
and the one remaining in the lower band is $1-P_{LZ}$, where $r$ is the ratio between the energy gap and the slope (in the standard
unit $\hbar\equiv 1$) given by 
\begin{align}\label{eq:characteristic_parameter}
    r\equiv(g/2)^2/s=\frac{3}{2}\frac{\delta\widetilde{k}^2_y}{\widetilde{E}},
\end{align}
which can be treated as a parameter characterizing the possible occurrence of
LZTs. Note that $P_{LZ}\approx4\%$ for $r=1$, so this provides a numerical 
criterion for determining if an LZT can occur: $0< r\leq 1$.

\begin{figure} [ht!]
\centering
\includegraphics[width=\linewidth]{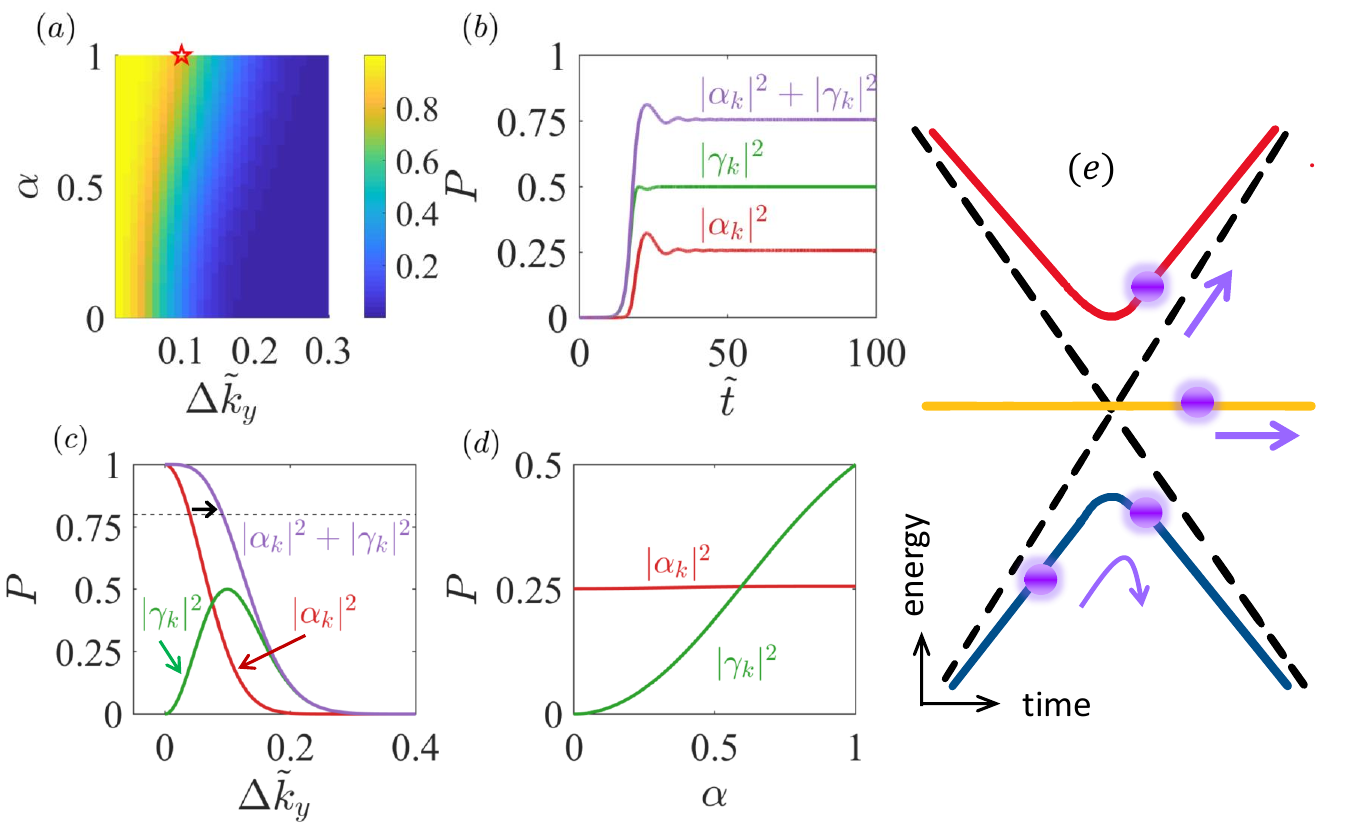}
\caption{LZT probabilities in the $\alpha$-$\mathcal{T}_3$ lattice. For electrons initiated from the lower band, there are two LZT probabilities:
$|\alpha_k|^2$ - the transition probability from the lower band to the upper
band, and $|\gamma_k|^2$ - the transition probability from the lower to the
flat bands.
(a) Color-coded sum of $|\alpha_k|^2$ and $|\gamma_k|^2$ in the 
$(\Delta \widetilde{k}_{y},\alpha)$ plane for $\widetilde{E}=0.0336$. The 
initial momenta are $\widetilde{k}_x=3$ and 
$\widetilde{k}_{y}=\Delta\widetilde{k}_{y}$ about $+\mathbf{K}$, and the 
integration time step is $d\widetilde{t}=0.01$. (b) Time evolution of the LZT 
probabilities for $\Delta\widetilde{k}_{y}$ and $\alpha$ values taken from the 
location of the pentagram in (a) with $|\alpha_{\pmb{k}}|^2+|\gamma_{\pmb{k}}|^2+|\beta_{\pmb{k}}|^2=1$. (c) LZT probabilities versus 
$\Delta\widetilde{k}_{y}$ for $\alpha=1$, where the arrow indicates the 
flat-band induced change in the LZT probability. (d) LZT probabilities versus 
$\alpha$ for $\Delta \widetilde{k}_{y}=0.1$. (e) LZT in the three-level system. If the initial state is in the lower band, after an LZT about the avoided crossing point, the state is a superposition of the eigenstates associated with all three bands.}
\label{fig:LZT_enhancement_flat}
\end{figure}

We next consider the opposite extreme case of the $\alpha$-$\mathcal{T}_3$ 
lattice: $\alpha = 1$ (pseudospin-1 lattice). Because of the presence of the 
flat band, the Hamiltonian linearized about a Dirac point can be related to 
that of the standard LZT model~\cite{carroll:1986} with three distinct energy 
levels. In particular, employing the unitary transformation
\begin{align}
    U=\exp\left(-\frac{i}{\hbar}\frac{\pi}{2}S_{y}\right)\exp\left(-\frac{i}{\hbar}\frac{\pi}{2}S_{z}\right),
\end{align}
we obtain the pseudospin-1 Hamiltonian as (Appendix~\ref{app:equivalence}) 
\begin{align}
     U^{\dagger}\widetilde{H}U\approx\frac{3\delta\widetilde{k}_y}{2}S_{x}\mp\frac{3\widetilde{E}\widetilde{t}'}{2}S_{z},
\end{align}
where $S_{x}$ and $S_{y}$ are the components of the vector of spin-1 matrices. 
The eigenenergy spectra of the upper and lower bands have the same form as 
Eq.~\eqref{eq:H_LZ_energy}, with the addition of the extra flat band in the 
middle of the lower and upper bands. For an electron initialized in the lower band, the LZT probabilities for it to
transition to the upper band, transition to the flat band and remain in the 
lower band are given by~\cite{carroll:1986} 
\begin{align}
    |\alpha_{\pmb{k}}|^2&=1-2\sqrt{P_{LZ}}+P_{LZ},\label{eq:LZT_three_level_up}\\
    |\gamma_{\pmb{k}}|^2& = 2\left(1-\sqrt{P_{LZ}}\right)\sqrt{P_{LZ}},\label{eq:LZT_three_level_flat}\\
    |\beta_{\pmb{k}}|^2&=P_{LZ},
\end{align}
respectively. 
where $P_{LZ}$ is defined in Eq.~\eqref{eq:formula_first_LZT_up}. To appreciate
the flat band contribution to the transitions, we set $P_{LZ}=1/4$ so that  
$|\alpha_{\pmb{k}}|^2=P_{LZ}$ holds for both two- and three-level systems
because $1-2\sqrt{P_{LZ}}=0$. Figure~\ref{fig:LZT_enhancement_flat}(d) shows 
the numerical result for $0\leq\alpha\leq 1$, where it can be seen that, in 
this special case, $|\alpha_{\pmb{k}}|^2$ is independent of the coupling 
strength between the flat and positive/negative bands.

We can now analyze the general $\alpha$-$\mathcal{T}_3$ lattice as an 
interpolation between the idealized two-level and three-level systems, where 
the coupling between the flat band and the other two bands varies in the range 
$0<\alpha<1$. For convenience, we again initialize electrons in the lower band 
and examine two transition probabilities: that from the lower to the upper band
denoted as $|\alpha_{\pmb{k}}|^2$ and that from the lower to the flat band 
denoted as $|\gamma_{\pmb{k}}|^2$. To unveil the effect of increasing the value
of $\alpha$ from zero, we calculate the two probabilities for $0\le\alpha\le 1$
and the range of momentum deviation $\Delta\widetilde{k}_{y}$ from the 
$+\mathbf{K}$ Dirac point in the hexagonal Brillouin zone.   
Figure~\ref{fig:LZT_enhancement_flat}(a) shows the color-coded sum of the two 
probabilities in the parameter plane $(\Delta\widetilde{k}_{y},\alpha)$, where 
the range of $\Delta\widetilde{k}_{y}$ to generate a high LZT probability 
increases monotonically with $\alpha$ and reaches maximum at $\alpha=1$, 
suggesting that the flat band enhances LZT. The time evolution of the two
probabilities and their sum is shown in Fig.~\ref{fig:LZT_enhancement_flat}(b).
Note that, for $\alpha = 1$, the LZT probabilities $|\alpha_{\pmb{k}}|^2$ and 
$|\gamma_{\pmb{k}}|^2$ can be determined from 
Eqs.~\eqref{eq:LZT_three_level_up} and ~\eqref{eq:LZT_three_level_flat},
respectively. Figure~\ref{fig:LZT_enhancement_flat}(c) shows, for $\alpha = 1$,
these two probabilities, together with their sum, versus 
$\Delta\widetilde{k}_{y}$, where the horizontal dashed line specifies $P=0.8$
and the arrow indicates the enhancement of the LZT by the flat band. As 
$\Delta\widetilde{k}_{y}$ increases, the value of the characteristic parameter 
$r$ increases, leading to a decrease in the LZT probability to the upper band.
However, even when the LZT probability from the lower to the upper bands is 
effectively zero, there can still be an appreciable transition probability from
the lower to the flat band. For example, for $\Delta\widetilde{k}_y=0.2$, we 
have $r\approx 1.8$. In this case, we have $|\alpha_{\pmb{k}}|^2 \approx 0$ but 
$|\gamma_{\pmb{k}}|^2 \approx 0.25$. Figure~\ref{fig:LZT_enhancement_flat}(d)
shows, for fixed $\Delta \widetilde{k}_{y}=0.1$, the two probabilities versus
$\alpha$. Note that the probability $|\gamma_{\pmb{k}}|^2$ increases 
monotonically with $\alpha$.

\begin{figure} [ht!]
\centering
\includegraphics[width=0.9\linewidth]{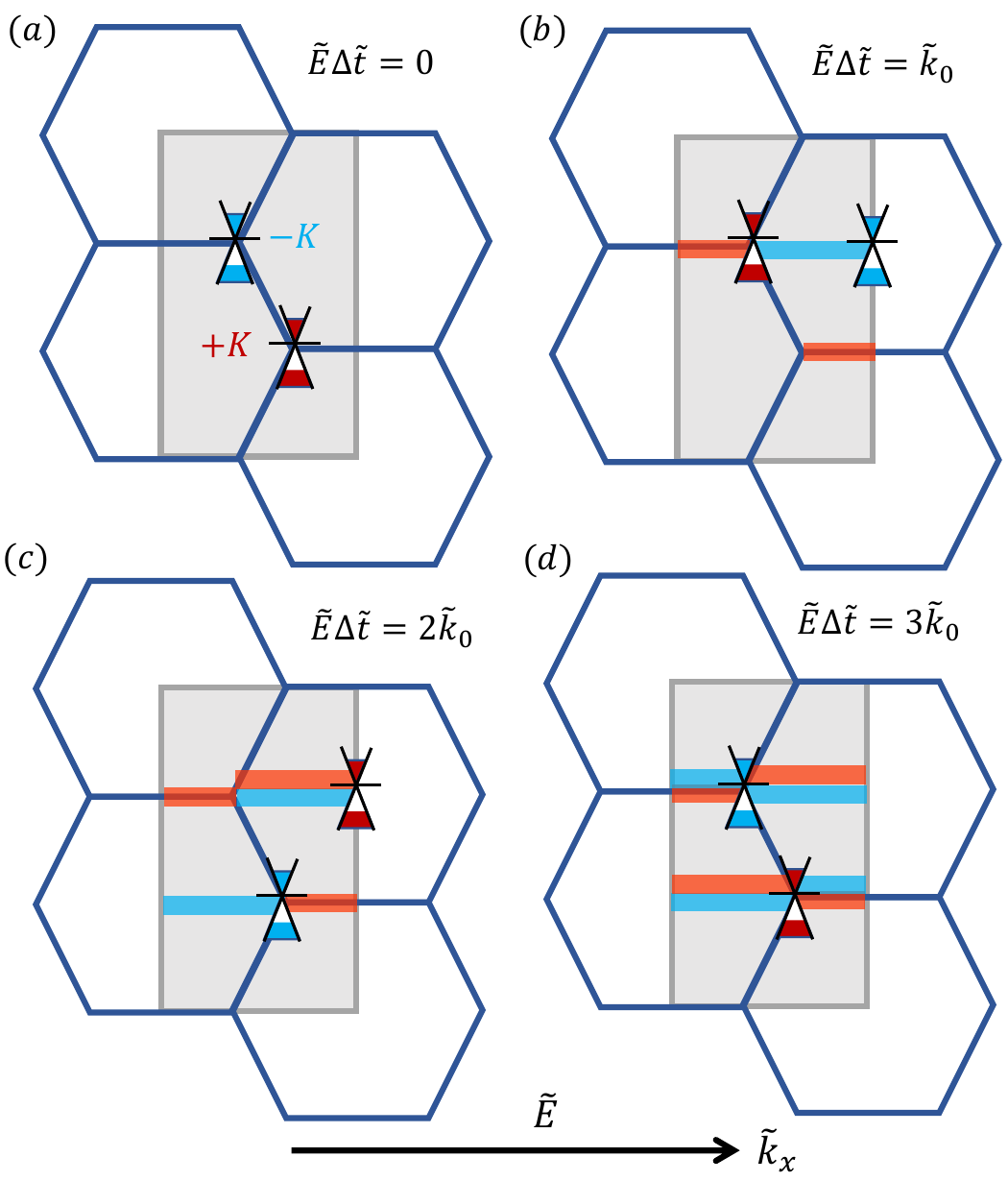}
\caption{Schematic illustration of Bloch-Zener oscillations. 
(a) Two nonequivalent Dirac points $\pm \mathbf{K}$ in the hexagonal Brillouin 
zone at $\widetilde{t}=0$, corresponding to two distinct valleys. Driven by a
static electric field in the positive $\widetilde{k}_x$ direction, both Dirac
points start to move in the same direction, where the two vertical sides of the 
gray rectangle denote the periodic boundaries in $\widetilde{k}_x$. When the electric field is applied to the $\alpha$-$\mathcal{T}_3$ lattice, in
the laboratory frame we have
$\widetilde{k}_{x}(t)=\widetilde{k}_{x}-\widetilde{E}\widetilde{t}$, for fixed
$\widetilde{k}_{x}$. In this ``static'' momentum space, an electron will move
toward the left with the momentum
$\widetilde{k}_{x}(t)\propto -\widetilde{E}\widetilde{t}$.
In the moving frame with $\widetilde{k}_{x}(t)$, the zero energy or Dirac
points will move toward the right with the momentum
$\widetilde{k}_{x}\propto \widetilde{E}\widetilde{t}$, as shown in Fig.~3 based
on the energy form in Eq.~\eqref{eq:energy}. Since LZTs occur around the Dirac 
points, it is convenient to follow the movements of the Dirac points. (b-d) The locations of $\pm \mathbf{K}$ after 
$\Delta\widetilde{t}=\widetilde{k}_{0}/\widetilde{E}$,
$2\widetilde{k}_{0}/\widetilde{E}$, and $3\widetilde{k}_{0}/\widetilde{E}$,
respectively. For $\Delta\widetilde{t}=3\widetilde{k}_{0}/\widetilde{E}$,
the two Dirac points return to their respective initial starting locations,
completing one cycle. The period of the Bloch-Zener oscillations is thus
$\widetilde{t}_{B}=3\widetilde{k}_{0}/\widetilde{E}$. During the Bloch
period, the Landau-Zener transition occurs twice about $\pm \mathbf{K}$, as 
indicated by the double horizontal color bars.}  
\label{fig:schematic_BZO}
\end{figure}

\begin{figure} [ht!]
\centering
\includegraphics[width=\linewidth]{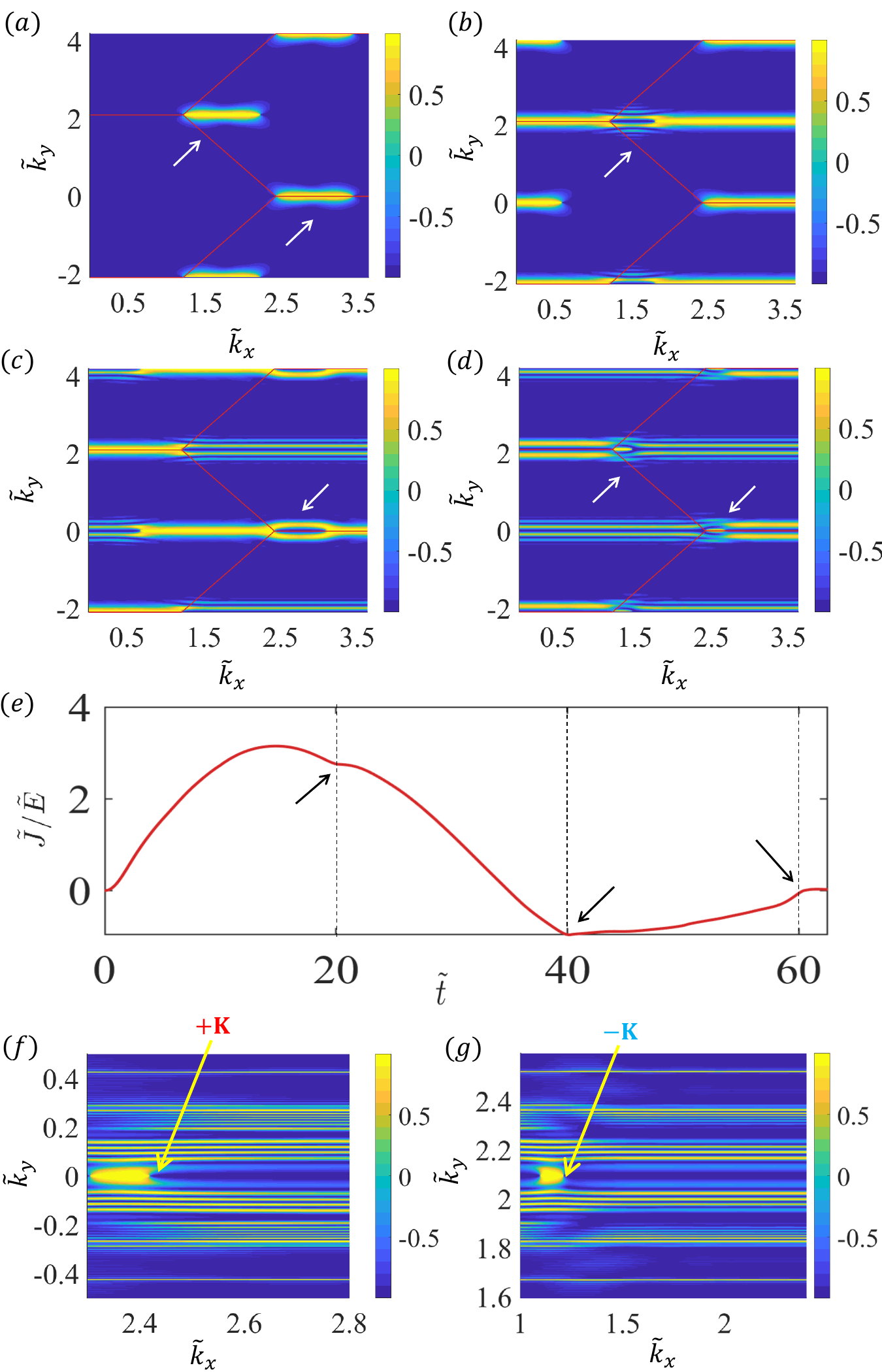}
\caption{Emergence of distinct LZT morphology and irregularities in the   
Bloch-Zener oscillations. (a) Morphology of LZT after $t = 0^+$ (immediately
after the electric field in the $\widetilde{k}_x$ direction is turned on).
Shown is the color-coded values of the differential LZT probability, defined as 
$\Delta P_{\alpha\beta}(t) \equiv \left|\alpha_{\pmb{k}}\left(t\right)\right|^{2}-\left|\beta_{\pmb{k}}\left(t\right)\right|^{2}$,
in the entire Brillouin zone. (b-d) Color-coded values of the differential
LZT probability $\Delta P_{\alpha\beta}(t)$ at three instants of time after 
which a morphological change in the LZTs occurs:   
$\Delta\widetilde{t}=\widetilde{k}_{0}/\widetilde{E}$,
$2\widetilde{k}_{0}/\widetilde{E}$ and $3\widetilde{k}_{0}/\widetilde{E}$.
(e) Evolution of the intraband current density within one Bloch period. At the 
three time instants indicated by the vertical dashed lines and arrows, the 
curve is nonsmooth, which correspond to the LZT morphology in (b-d), 
respectively, and signify irregularities in the Bloch-Zener oscillations. 
(f,g) Morphology of LZT after 20 periods of Bloch-Zener oscillations with 
magnification about +$\mathbf{K}$, -$\mathbf{K}$, respectively. Simulation 
parameter values are $\alpha=0$, $\widetilde{E}=0.1200$, 
$d\widetilde{t}\approx0.01$, and 
$d\widetilde{k}_x\approx d\widetilde{k}_y\approx0.012$.}
\label{fig:irregular_BZO}
\end{figure}

\subsection{Morphology and irregular Bloch-Zener oscillations}\label{subsec:morphology_irregularBZO}

Under a static electric field, the intraband current density will exhibit 
periodic-like Bloch-Zener oscillations, where the interband contribution can 
be neglected in the long time situation~\cite{dora:2010}. 
Figure~\ref{fig:schematic_BZO} provides a schematic picture to explain the 
origin of the oscillations. Driven by a constant electric field in the positive
$\widetilde{k}_x$ direction, after $\widetilde{t}=0^+$, the two Dirac points 
$\pm \mathbf{K}$ in the hexagonal Brillouin zone start to move in the same 
direction, where the gray rectangular region denotes a periodic area in the 
momentum space, as shown in Fig.~\ref{fig:schematic_BZO}(a). The edge length 
of the hexagonal Brillouin zone is $\widetilde{k}_{0}=4\pi/(3\sqrt{3})$ in 
units of $1/a$, the inverse of the lattice constant. After the time 
$\Delta\widetilde{t}=\widetilde{k}_{0}/\widetilde{E}$, the $+\mathbf{K}$ valley
reaches the original location of the $-\mathbf{K}$ valley, as shown in 
Fig.~\ref{fig:schematic_BZO}(b). The $-\mathbf{K}$ reaches the original 
location of the $+\mathbf{K}$ after the time 
$\Delta\widetilde{t}=2\widetilde{k}_{0}/\widetilde{E}$, as shown in 
Fig.~\ref{fig:schematic_BZO}(c). The occurrences of the LZTs associated with 
the $\pm\mathbf{K}$ valleys are indicated by the orange and blue horizontal
strips, respectively, where the width of excitation zone depends on both the
gap size and the magnitude of electric field. At the time
$\widetilde{t}_{B}=3\widetilde{k}_{0}/\widetilde{E}$, both Dirac points 
return to their original starting locations, completing one cycle of 
oscillation during which LZT occurs twice. 

In the vicinity of a Dirac point ($+\mathbf{K}$ or $-\mathbf{K}$) in the 
momentum space, an infinite set of energy gaps exists where, for different 
momentum deviations $\Delta\widetilde{k}_y$ from the Dirac point, the sizes 
of the energy gap can be quite distinct from the exact energy gap in 
Eq.~(\ref{eq:gap_delta_k_y}). As the static electric field in the $x$ 
direction is turned on, the momentum in the $+\widetilde{k}_x$ direction 
increases linearly with time, sweeping through all possible values of the $x$ 
component of the momentum in the Brillouin zone, effectively eliminating all 
the original differences in $\widetilde{k}_x$ for different points in the 
momentum space. However, the various deviations in the $y$ component of the 
momentum, i.e., the different $\Delta\widetilde{k}_y$ values, still matter and 
in fact persist because they correspond to different energy gaps. As a result, 
different values of $\Delta\widetilde{k}_y$ will lead to different 
probabilities of LZT, creating a distinct morphology with respect to the LZT 
probability in the $\widetilde{k}_y$ direction of the momentum space at any 
given time. As an example, Figs.~\ref{fig:irregular_BZO}(f) and 
\ref{fig:irregular_BZO}(g) present such a morphology after 20 periods of 
Bloch-Zener oscillations with the magnification about the $\pm\mathbf{K}$
Dirac points, respectively, where the color-coded values of the differential 
LZT probability 
$\Delta P_{\alpha\beta}(t) \equiv \left|\alpha_{\pmb{k}}\left(t\right)\right|^{2}-\left|\beta_{\pmb{k}}\left(t\right)\right|^{2}$
are shown in the $(\widetilde{k}_x,\widetilde{k}_y)$ plane.
An observation is that the LZT morphology, as exemplified in 
Figs.~\ref{fig:irregular_BZO}(f) and \ref{fig:irregular_BZO}(g), can undergo 
changes due to the complicated interference pattern between quantum states in 
different energy bands in the long time, due to the non-deterministic nature 
of the LZTs with respect to the size of the energy gap. 
Figure~\ref{fig:irregular_BZO}(a) presents such a morphology after time $t=0^+$
(immediately after the electric field is turned on) within one period of the 
Bloch-Zener oscillation. Three examples are illustrated in 
Fig.~\ref{fig:irregular_BZO}(b-d), where the color-coded values of 
$\Delta P_{\alpha\beta}(t)$ at three time instants:    
$\Delta\widetilde{t}=\widetilde{k}_{0}/\widetilde{E} \approx 20$, 
$2\widetilde{k}_{0}/\widetilde{E} \approx 40$ and 
$3\widetilde{k}_{0}/\widetilde{E} \approx 60$, are displayed. The morphological
changes in the LZTs are concentrated in the vertical neighborhoods of the Dirac
points, whereas the values of $\Delta P_{\alpha\beta}(t)$ in most of the 
momentum space remain unchanged.

A remarkable phenomenon is that, when a change in 
the LZT morphology occurs, some experimentally measurable quantities such as 
the current density can undergo a sudden change as well. To be concrete, we
focus on the current density associated with Bloch-Zener oscillations, which 
is dominantly determined by the intraband behaviors~\cite{dora:2010}. Suppose
that, initially, the electrons are prepared in the lower band. Due to the 
change in the LZT morphology and the mutual switching between the Dirac points 
$\pm\mathbf{K}$, the dependence of the probabilities for the electrons to be 
in the upper band on the momentum will change, and this will lead to a sudden 
change in the current density that is contributed to by all the momenta in the 
Brillouin zone. Note from Figs.~\ref{fig:irregular_BZO}(b-d) that the changes 
in the LZT morphology are pronounced only near the original Dirac points 
where an LZT is most likely to occur, while there are no such changes for most 
of the momentum space. Since the current density is the integration over the 
entire Brillouin zone, the resulting change in the current density will be 
quite ``subtle'' in the sense that it will not be a discontinuous change in 
the current density itself but a non-smooth change (or, equivalently, a 
discontinuous change in the time derivative of the current density). Such 
nonsmooth changes have indeed been numerically observed, as shown in 
Fig.~\ref{fig:irregular_BZO}(a), where the arrows and the vertical dashed lines
indicate the three time instants at which such a change occurs, corresponding 
to the distinct LZT morphology in Figs.~\ref{fig:irregular_BZO}(b-d), 
respectively. To compare irregular and periodic Bloch-Zener oscillations, we 
also study the oscillations resulting from near ideal LZTs in the momentum 
space (Appendix~\ref{app:reason_irregular}).

\section{Landau-Zener-St\"{u}ckelberg interferometry in $\alpha$-$\mathcal{T}_3$ lattice} \label{sec:LZSI}

\subsection{Two- and three-level models}\label{subsec:two_three_LZSI}

The dynamical evolution of the wavefunction in the $\alpha$-$\mathcal{T}_3$ 
lattice for $\alpha=0$ ($1$) driven by a static electric field can be 
described by the Hamiltonian of a strongly periodically driven two- or 
three-level system, as demonstrated in Appendix~\ref{app:equivalence}. 
According to the adiabatic-impulse theory~\cite{Damski:2006,shevchenko:2010}, 
the quantum evolution can be decomposed into adiabatic evolution and 
non-adiabatic LZTs, where the former occurs most of the time but 
the latter occur on a short time scale. For $0 < \alpha < 1$, under the 
adiabatic impulse approximation, this physical picture still applies. 
Specifically, for the adiabatic evolution, the eigenenergy spectrum is 
independent of the value of the lattice coupling parameter $\alpha$, so the 
adiabatic phase for $0<\alpha<1$ is similar to that for $\alpha=1$. Insights 
into the LZTs can be gained by numerically calculating the time evolution of 
the transition probability $|\alpha_k|^2$ and $|\gamma_k|^2$ in 
$\alpha$-$\mathcal{T}_3$ lattice, as exemplified in 
Fig.~\ref{fig:LZT_enhancement_flat}. 

\begin{figure} [ht!]
\centering
\includegraphics[width=\linewidth]{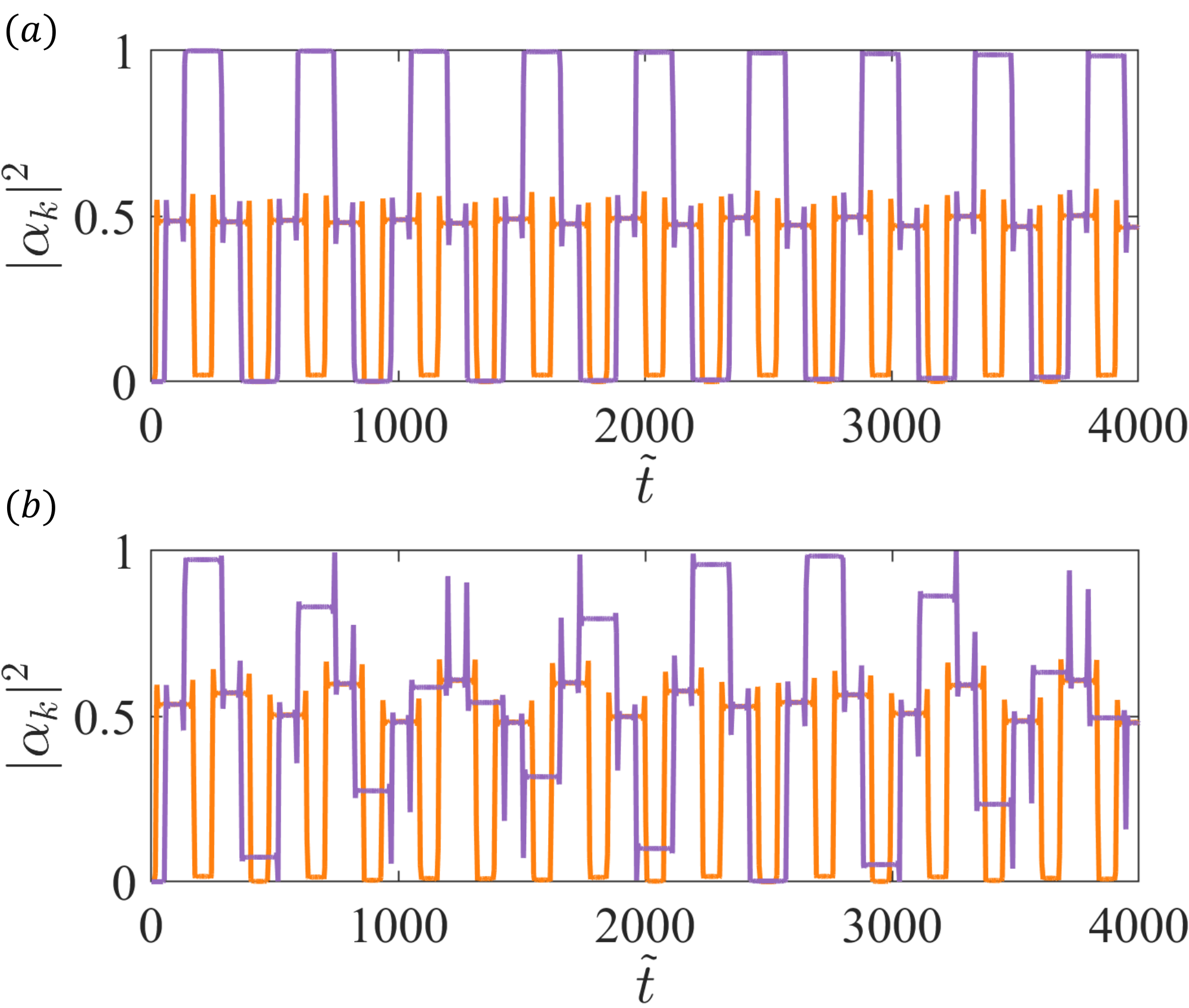}
\caption{Landau-Zener transitions in $\alpha$-$\mathcal{T}_3$ lattice for 
$\alpha=0$. Shown in (a,b) is the time evolution of $|\alpha_{\pmb{k}}|^2$, 
the probability of LZT from the lower to the upper band, for four different 
initial momentum values. The orange traces correspond to the case where the
momentum is initiated in the vicinity of the $+\mathbf{K}$ valley and within 
one Bloch period, LZT is required for $|\alpha_{\pmb{k}}|^2$ to reach zero. 
The purple traces are for the case where the momentum is initiated in the 
vicinity of the $-\mathbf{K}$ valley. In (a), the initial momentum values
for the orange and purple traces are $\widetilde{k}_y=0.07$ about 
$+\mathbf{K}$ and $2\pi/3+0.07$ around $-\mathbf{K}$, respectively. In (b), the 
corresponding orange and purple traces are the results of setting the initial 
momentum values to $\widetilde{k}_y=0.065$ (about $+\mathbf{K}$) and 
$2\pi/3+0.065$ (around $-\mathbf{K}$), respectively. The transition behaviors
displayed can be understood in terms of the St\"{u}ckelberg phase (to be 
analyzed below) that tends to be approximately constant when the initial 
momentum is near the $+\mathbf{K}$ valley but exhibits large variations when 
the initial momentum is near the $-\mathbf{K}$ valley. Other parameter values 
are $\widetilde{E}=0.0317$, $\widetilde{k}_{x} = 3$ and $d\widetilde{t}=0.01$.}
\label{fig:noflat_double_passage_like_complete}
\end{figure}

The adiabatic evolution of the wavefunction generates an adiabatic phase, while
the LZTs lead to a non-adiabatic phase. To gain insights, consider 
the double passage case in a strongly periodically driven two-level system, 
where two successive transitions are required for a particle initiated from an 
eigenstate in the lower band to reach the upper band with probability one, as 
exemplified in Fig.~\ref{fig:noflat_double_passage_like_complete}(a). In this 
case, after two successive LZTs, the transition probability to the upper band 
is given by~\cite{liang:2020,shevchenko:2010} 
(Appendix~\ref{subapp:adiabatic_impulse_two_level})
\begin{align}\label{eq:double_passage}
    P_{+} &= 4P_{LZ}(1-P_{LZ})\sin^2(\Phi_{st}),\\
    \Phi_{st} &= \zeta + \varphi_s,
\end{align}
where $P_{LZ}$ is the first-time LZT probability given by 
Eq.~\eqref{eq:formula_first_LZT_up} and $\Phi_{st}$ is the St\"{u}ckelberg 
that consists of two components: the adiabatic phase $\zeta$ between two 
consecutive LZTs and the non-adiabatic phase, i.e., the Stokes phase 
$\varphi_S$ at the transition. There are two distinct cases. The first is  
\begin{align}
\Phi_{st}=\pi/2 + k\pi,\;k\in\mathbf{Z}, 
\end{align}
corresponding to constructive interference~\cite{shevchenko:2010}
because, after one driving period, the maximum transition probability to the 
upper band is $P_{+}=4P_{LZ}(1-P_{LZ})$, which is twice the average transition 
probability $\langle{P}_{+}\rangle=2P_{LZ}(1-P_{LZ})$ over one period. The 
second case is 
\begin{equation}
    \Phi_{st}= k\pi,
\end{equation} 
giving rise to destructive interference~\cite{shevchenko:2010} as $P_{+}=0$ 
after one driving period.

Our equivalence analysis in Appendix~\ref{app:equivalence} and the treatment 
of the strongly periodically driven two-level system in 
Appendix~\ref{subapp:adiabatic_impulse_two_level} give that, for $\alpha = 0$
in the $\alpha$-$\mathcal{T}_3$ lattice, the adiabatic phase is given by
\begin{align} \label{eq:adiabatic_phase}
    \zeta=\int\text{\ensuremath{\epsilon_{\pmb{k}}(t)dt}},
\end{align}
where $\epsilon_{\pmb{k}}(t)$ is defined in Eq.~\eqref{eq:energy} and 
depends on the electric field $\widetilde{E}$ and the momentum deviation 
$\Delta\widetilde{k}_{y} $ from $\pm \mathbf{K}$. The non-adiabatic phase 
$\varphi_{s}$ is determined by the linearized LZT Hamiltonian about the
Dirac points $\pm \mathbf{K}$:
\begin{align}\label{eq:stocks_phase}
   \varphi_{s}=\pi/4+\delta(\ln\delta-1)+\arg\Gamma(1-i\delta),
\end{align}
where $\delta$ is determined by the LZT characteristic parameter as 
$2\delta=r$, as given by Eq.~\eqref{eq:characteristic_parameter}. It can be
seen that the St\"{u}ckelberg phase also depends on the electric field 
$\widetilde{E}$ and the momentum deviation $\Delta \widetilde{k}_{y}$ from 
$\pm \mathbf{K}$. 

\begin{figure} [ht!]
\centering
\includegraphics[width=\linewidth]{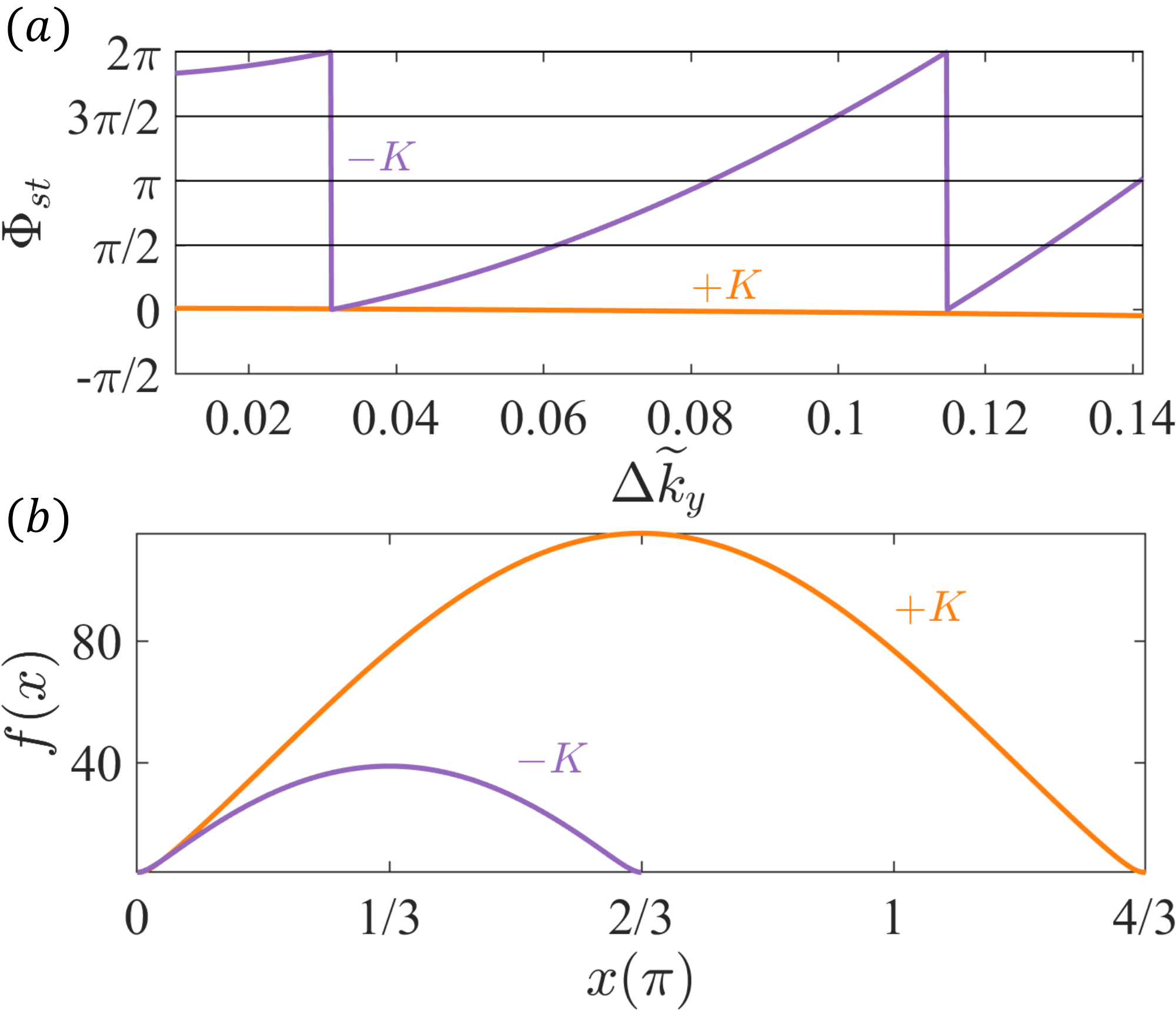}
\caption{St\"{u}ckelberg phase in $\alpha$-$\mathcal{T}_3$ lattice for 
$\alpha = 0$. (a) The St\"{u}ckelberg phase versus the momentum deviation 
from $+\mathbf{K}$ (orange) and $-\mathbf{K}$ (purple) valleys. The range of 
the momentum deviation is determined by the empirical criterion: $0<r\leq1$, 
which is $\Delta\widetilde{k}_y\in[0.01,(2\widetilde{E}/3)^{1/2}]$ for 
$\widetilde{E}=0.03$ and $d\Delta\widetilde{k}_y=0.0001$. The integration time 
step is $d\widetilde{t}=0.01$. (b) The adiabatic phase measured from the Dirac 
points $\pm\mathbf{K}$ as determined by the integral of function $f(x)$ over 
$x\in[0,2\pi/3]$($-\mathbf{K}$), $x\in[0,4\pi/3]$($+\mathbf{K}$) with 
$\Delta\widetilde{k}_y=0.7$.}
\label{fig:stuckelberg phase}
\end{figure}

Figure~\ref{fig:stuckelberg phase}(a) shows the St\"{u}ckelberg phase for two
types of initial momentum values for $\widetilde{E}=0.03$. For LZTs starting 
near the Dirac point $+\mathbf{K}$, the St\"{u}ckelberg phase is non-adiabatic
and approximately independent of the momentum deviation 
$\Delta\widetilde{k}_{y}$, while for $-\mathbf{K}$, the phase is adiabatic so 
it depends on the momentum deviation [Eq.~(\ref{eq:stocks_phase})]. More
specifically, for LZTs starting from the Dirac points $\pm\mathbf{K}$, the 
adiabatic phase between two successive LZTs is
\begin{align}\label{eq:adiabaticPhase}
    \zeta_{+\mathbf{K}}&=\int^{\frac{4\pi}{3}}_{0} f_{+\mathbf{K}}(x)dx,\\    \zeta_{-\mathbf{K}}&=\int^{\frac{2\pi}{3}}_{0} f_{-\mathbf{K}}(x)dx,
\end{align}
where the functions $f_{+\mathbf{K}}(x)$ and $f_{-\mathbf{K}}(x)$ are given by
\begin{align}\label{eq:f_pm}
    f_{+\mathbf{K}}&=\frac{2}{\sqrt{3}\widetilde{E}}\sqrt{1+4a_{+}\cos\left(\frac{2\pi}{3}-x\right)+4\cos^2\left(\frac{2\pi}{3}-x\right)},\nonumber\\
    f_{-\mathbf{K}}&=\frac{2}{\sqrt{3}\widetilde{E}}\sqrt{1+4a_{-}\cos\left(\frac{\pi}{3}-x\right)+4\cos^2\left(\frac{\pi}{3}-x\right)}
\end{align}
with $a_{+}\equiv\cos(3\Delta\widetilde{k}_y/2)$, $a_{-}=-a_{+}$, 
$\Delta\widetilde{k}_y$ measured from $\pm\mathbf{K}$, and 
$x\equiv\widetilde{E}\widetilde{t}\sqrt{3}/2$. 
Figure~\ref{fig:stuckelberg phase}(b) shows the adiabatic phase measured from 
the Dirac points $\pm\mathbf{K}$. Since the integration of $f_{+\mathbf{K}}$ 
has a relatively large value, the adiabatic phase $\zeta_{+\mathbf{K}}$ is 
nearly constant for different momentum deviation $\Delta\widetilde{k}_y$ 
for $0<r\leq1$, whereas $\zeta_{-\mathbf{K}}$ is sensitive to 
$\Delta\widetilde{k}_y$. The adiabatic phase also depends on the 
electric field $\widetilde{E}$. (More elaborate details can be found in
Appendix~\ref{app:adiabatic_electric}.)

\begin{figure} [ht!]
\centering
\includegraphics[width=1\linewidth]{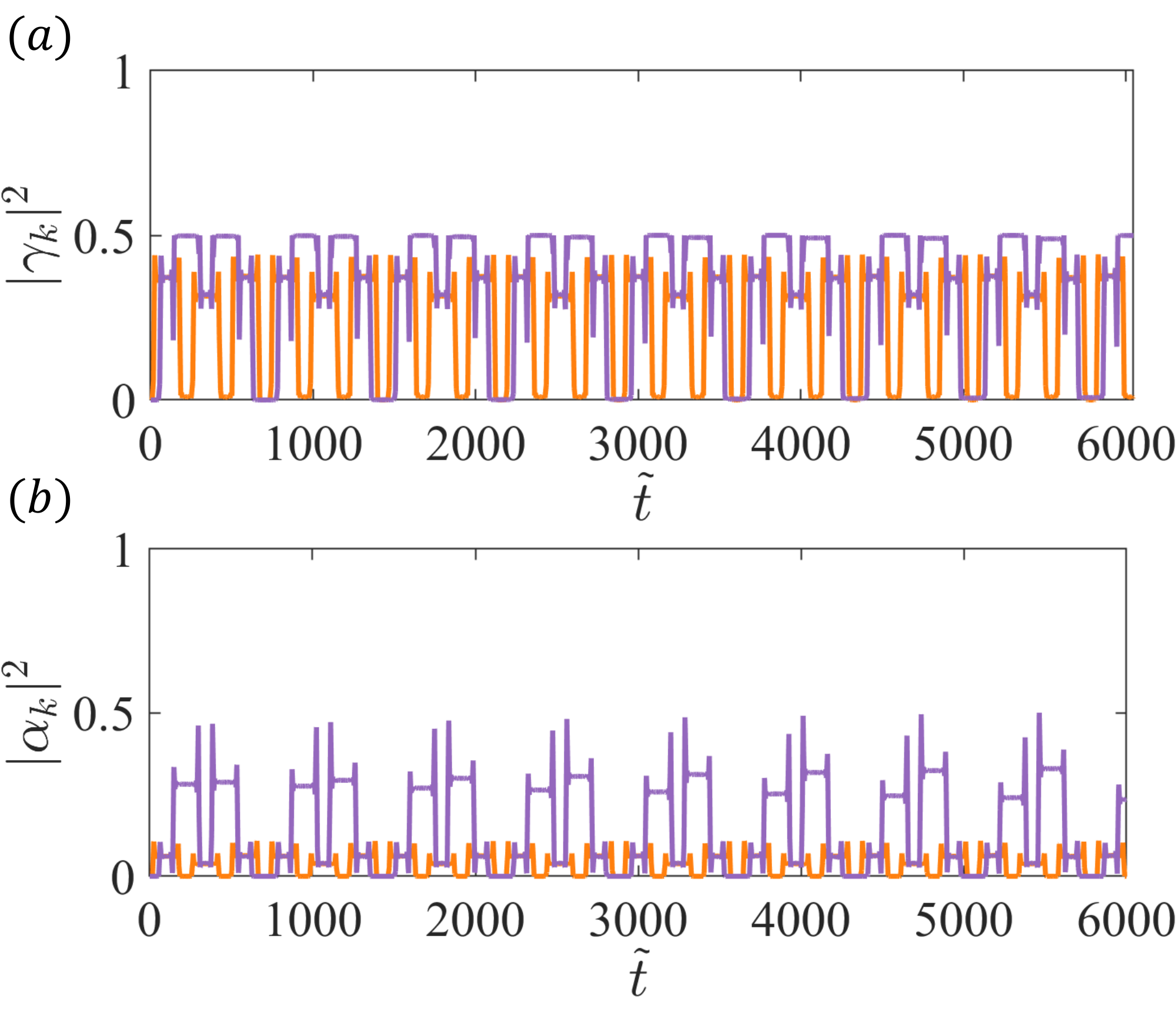}
\caption{Landau-Zener transitions in $\alpha$-$\mathcal{T}_3$ lattice for
$\alpha=1$. Because of the presence of a flat band, the time evolution of 
two quantities is displayed: (a) $|\gamma_{\pmb{k}}|^2$ and 
(b) $|\alpha_{\pmb{k}}|^2$, the transition probability from the flat to the 
upper band and that from the lower to the upper band, respectively. 
For the orange traces, the initial momentum is $\widetilde{k}_{y}=0.135$ 
around $+\mathbf{K}$. For the purple traces, the initial momentum is 
$\widetilde{k}_{y}=2\pi/3+0.135$ around $-\mathbf{K}$. Other parameters are 
$\widetilde{E}=0.03$, $\widetilde{k}_x=3$, and $d\widetilde{t}=0.01$.} 
\label{fig:flat_double_passage_like_complete}
\end{figure}

\begin{figure*} [ht!]
\centering
\includegraphics[width=0.8\linewidth]{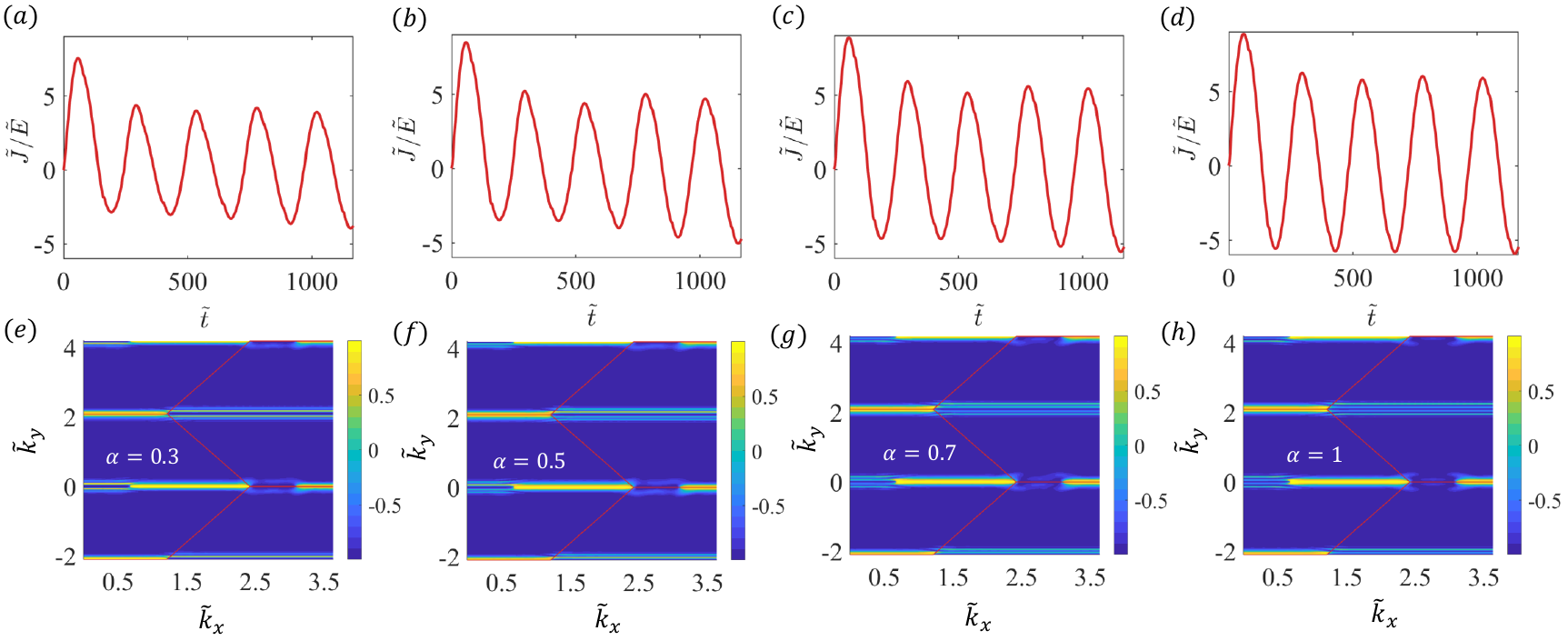}
\caption{Mostly regular (or weakly irregular) Bloch-Zener oscillations in 
general $\alpha$-$\mathcal{T}_3$ lattices. (a-d) Oscillations of the intraband
current density for $\alpha = 0.3$, 0.5, 0.7, and 1.0, respectively. 
(e-h) The corresponding LZT morphology revealed by the color-coded 
differential LZT probability $\Delta P_{\alpha\beta}(t)$ in the momentum space 
at a specific time, where the St\"{u}ckelberg phase $\Phi_{st}=0$ for LZTs 
starts from $+\mathbf{K}$ valley. Simulation parameter values are 
$\widetilde{E}=0.03$, $d\widetilde{t}\approx0.001$, and 
$d\widetilde{k}_x \approx d\widetilde{k}_y \approx 0.012$.}
\label{fig:flat_retention_SBZO}
\end{figure*}

\begin{figure*} [ht!]
\centering
\includegraphics[width=0.8\linewidth]{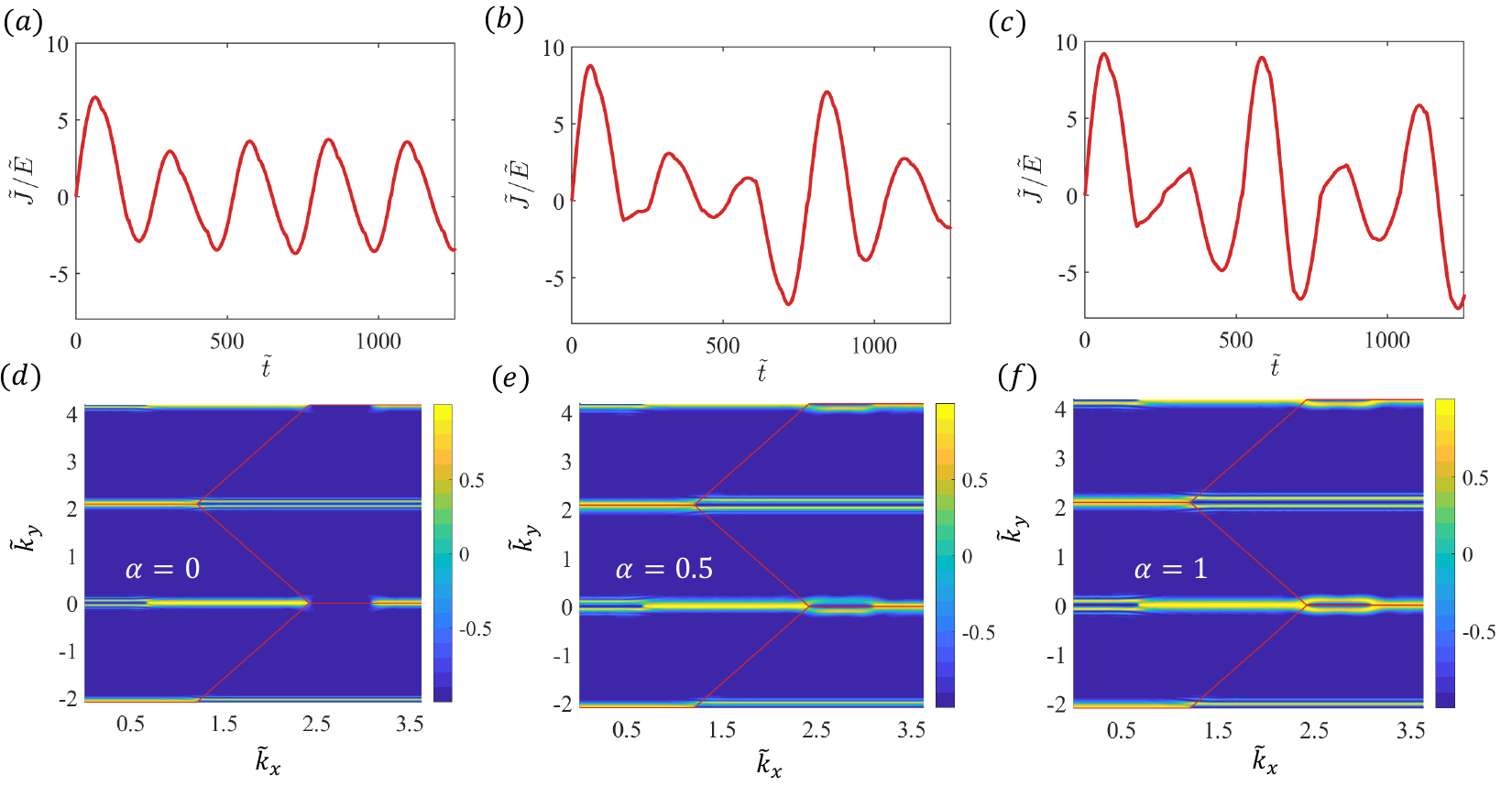}
\caption{Strongly irregular (``chaotic'') Bloch-Zener oscillations. The static
electric field strength is $\widetilde{E}=0.0279$. (a) Mostly regular 
oscillations in the intraband current density for $\alpha = 0$ (graphene).
As $\alpha$ increases from zero, the oscillations become strongly irregular,
as shown in (b) and (c) for $\alpha =0.5$ and $\alpha = 1$, respectively.
(d-f) The corresponding LZT morphology at a particular time instant, 
where the St\"{u}ckelberg phase $\Phi_{st}=\pi$ for LZTs starts from
the $+\mathbf{K}$ valley. Integration parameter values are 
$d\widetilde{t}\approx0.01$ and 
$d\widetilde{k}_x\approx d\widetilde{k}_y\approx0.012$.}
\label{fig:flat_destroy_SBZO}
\end{figure*}

For $\alpha > 0$, a flat band arises in addition to the two Dirac cone bands.
Figure~\ref{fig:flat_double_passage_like_complete} shows the representative 
time evolution of the transition probabilities $|\alpha_k|^2$ and 
$|\gamma_k|^2$ for $\alpha = 1$. Exploiting the equivalence of the 
dice lattice driven by a constant electric field to a strongly periodically 
driven three-level system (Appendix~\ref{app:equivalence}), we have that, after
one Bloch-Zener oscillation period (two successive LZTs), the occupation 
probability of the upper, flat and lower bands are given by 
(Appendix~\ref{subapp:adiabatic_impulse_three_level})
\begin{align}
    &P_{+}=16\widetilde{P}_{LZ}^2\sin^{4}(\zeta/2),\label{eq:main_body_three_level_upper_phase}\\
    &P_{0}=2\widetilde{P}_{LZ}((1-4\widetilde{P}_{LZ})(1-\cos\zeta)^2+\sin^2\zeta),\label{eq:main_body_three_level_flat_phase}\\
    &P_{-}=(2\widetilde{P}_{LZ}\cos\zeta+(1-2\widetilde{P}_{LZ}))^2 \label{eq:main_body_three_level_lower_phase}
\end{align}
with the normalization constraint $P_{+}+P_{0}+P_{-}=1$, where
\begin{align} 
\widetilde{P}_{LZ}\equiv P_{LZ}^{1/2}(1-P_{LZ}^{1/2}). 
\end{align}
For the upper band, $\zeta=\pi+2k\pi,\;k\in \mathbf{Z}$ corresponds to 
constructive interference and $\zeta=2k\pi$ leads to destructive 
interference [Eq.~\eqref{eq:main_body_three_level_upper_phase}]. For the flat 
band, $\zeta=2k\pi$ gives $P_0 = 0$ 
[Eq.~\eqref{eq:main_body_three_level_flat_phase}]. For $\zeta=\pi+2k\pi$, 
we have $P_0 \ne 0$ with $\widetilde{P}_{LZ}=1/4$. Note that the Stokes phase 
disappears in this case based on the non-adiabatic transition 
matrix~\cite{carroll:1986}. From Eq.~\eqref{eq:stocks_phase}, for 
$0<r=2\delta<1$, the Stokes phase is a small constant: $\varphi_s\approx0.5$. 
As a result, we have $\Phi_{st}\approx\zeta$. 

\subsection{Morphology selection}\label{subsec:design_morphology}

So far, we have numerically observed the complex LZT morphology and found 
a relationship between morphological changes and irregular Bloch-Zener 
oscillations. In addition, we have analyzed the interference phases in 
$\alpha$-$\mathcal{T}_3$ lattice based on the equivalence between the 
lattice subject to a constant electric field and strongly periodically 
driven two- or three level systems. We have found that the St\"{u}ckelberg 
phase (adiabatic phase) of LZTs starting from the Dirac point $+\mathbf{K}$ 
is nearly independent of the momentum deviation $\Delta\widetilde{k}_y$, i.e., 
the energy gap, while the phase starting from $-\mathbf{K}$ is sensitive to 
$\Delta\widetilde{k}_y$. For $\Phi_{st}=0,\pi$, destructive interference 
among the quantum states from different bands arises for $\alpha=0$. For
$\alpha=1$, only $\Phi_{st}=0$ corresponds to destructive interference. 
These findings suggest a principle of morphology selection for 
$0\leq\alpha\leq1$: setting $\Phi_{st}=0$ or $\Phi_{st}=\pi$ for LZT starting 
from the $+\mathbf{K}$ valley will result in two distinct types of LZT 
morphology, as illustrated in Figs.~\ref{fig:flat_retention_SBZO} and 
\ref{fig:flat_destroy_SBZO}, respectively. As analyzed in 
Appendix~\ref{app:reason_irregular}, ideal LZTs in the Brillouin zone lead to 
periodic and regular Bloch-Zener oscillations. The destructive interference 
has a similar effect to that of ideal LZTs. AS a result, the resulting 
morphology pattern can improve the regularity of Bloch-Zener oscillations. 

Figure~\ref{fig:flat_retention_SBZO} shows, for $\widetilde{E}=0.03$ and 
four different types of $\alpha$-$\mathcal{T}_3$ lattices, the irregular
Bloch-Zener oscillations and the corresponding representative LZT morphology at
a given time in the momentum space. It can be seen that the Bloch-Zener 
oscillations shown in Figs.~\ref{fig:flat_retention_SBZO}(a-d) are somewhat 
irregular - they are mostly regular. The destructive interference pattern of 
LZTs starting from the $+\mathbf{K}$ valley in the momentum space is shown in 
Figs.~\ref{fig:flat_retention_SBZO}(e-h) with $\Phi_{st}=0$. For another
type of morphology with $\Phi_{st}=\pi$, the Bloch-Zener oscillations are more 
irregular. For example, for $\alpha \ne 0$, the oscillations can be strongly 
irregular or ``chaotic,'' as exemplified in 
Figs.~\ref{fig:flat_destroy_SBZO}(b) and \ref{fig:flat_destroy_SBZO}(c). 
In this case, for $0\le \alpha\le 1$, the interference pattern associated
with the LZTs starting from the $+\mathbf{K}$ valley changes from destructive 
to non-destructive, as shown in Figs.~\ref{fig:flat_destroy_SBZO}(d-f). In 
both cases, the morphology of LZTs starting from the $-\mathbf{K}$ valley 
changes little due to the sensitivity to the momentum deviation. 

\section{Discussion} \label{sec:discussion}

Bloch or Bloch-Zener oscillations, in addition to being a fundamental 
phenomenon in solid state physics, practically provides the foundation to 
convert a direct current into an oscillating current in the terahertz frequency 
regime~\cite{schubert:2014}. Research in the past decade or so has suggested 
that, in 2D multiband materials, Bloch-Zener oscillations can be vulnerable to 
Landau-Zener transition or 
tunneling~\cite{breid:2006,breid:2007,kling:2010,lim:2015,sun:2018,chang:2022}.
Is this generally true? The question is important to physics and our present 
work has provided two-fold answers: yes LZTs can indeed affect the Bloch 
oscillations but what is destroyed is not the oscillations themselves but
just the perfect {\em time periodicity} of the oscillations; no because the 
irregular oscillations can persist even with frequent occurrences of LZTs. In 
fact, approximately periodic Bloch-Zener oscillations can be maintained if the
LZTs are near ideal such that they result in a near-one probability for the 
electrons to switch into a different band or when near destructive interference
arises between the quantum states in different energy bands. Nonsmooth or 
irregular behaviors in the current density arise when the LZTs are not
ideal and the interference is partially destructive. 

Our study encompasses the entire spectrum of a class of 2D Dirac materials 
modeled by $\alpha$-$\mathcal{T}_3$ lattices. We have found that the set of 
points in the 2D momentum space near a Dirac point at which LZTs occur can 
possess a complex morphology, and it is the change in the morphology that 
results in irregular Bloch-Zener oscillations. Theoretically, when driven by 
a static electric field, an $\alpha$-$\mathcal{T}_3$ lattice is equivalent 
the Landau-Zener-St\"{u}ckelberg interferometry. Specifically, the 
$\alpha = 0$ lattice (graphene) is effectively a two-level periodically 
driven quantum system while the general $\alpha$-$\mathcal{T}_3$ lattice for 
$0 < \alpha \le 1$ is equivalent to a three-level periodically driven system. 
For the three-level system, we have exploited the concept of St\"{u}ckelberg 
phase from the adiabatic impulse theory to understand the LZTs that occur in 
the neighborhoods of the Dirac points in distinct valleys. 

In the $\alpha$-$\mathcal{T}_3$ lattice, a non-zero coupling parameter $\alpha$
induces a flat band in between the positive and negative energy bands,
effectively resulting a three-level system. After the first LZT in 
Fig.~\ref{fig:LZT_enhancement_flat}, the quantum state is a superposition of 
the three energy bands and the LZT is enhanced by the flat band (compared with 
the two-level case). For subsequent LZTs (e.g., as shown in 
Figs.~\ref{fig:flat_retention_SBZO} and \ref{fig:flat_destroy_SBZO}), the flat 
band modifies the St\"{u}ckelberg phase $\Phi_{st}$ for destructive 
interference in the three-band interferometry (compared with the two-band one).
More specifically, for $\alpha=0$, the destructive interference corresponds to 
$\Phi_{st}=0,~\pi$ from Eq.~\eqref{eq:double_passage}. For $\alpha=1$, the 
destructive interference means $\Phi_{st}=0$ only and the case 
$\Phi_{st}=\pi$ is excluded from 
Eqs.~(\ref{eq:main_body_three_level_upper_phase}-\ref{eq:main_body_three_level_lower_phase}). For $0<\alpha<1$, the behavior of destructive interference is 
obtained numerically, as shown in Figs.~\ref{fig:flat_retention_SBZO} and 
\ref{fig:flat_destroy_SBZO}, which is slightly different from the $\alpha=1$ 
case, especially in Figs.~\ref{fig:flat_destroy_SBZO}(e,f). Taken together, a
nonzero $\alpha$ modifies the physical picture from two- to three-band
quantum interference: it generates a flat band in the original two-level
system, creates three-band interferometry, enhances LZT in the momentum space
around the Dirac points, and modifies destructive interference.

Theoretically, the asymmetrical morphology pattern around the $\pm K$ points
results from the different characteristics of the St\"{u}ckelberg phases
$\Phi_{st}$. As shown in Fig.~\ref{fig:stuckelberg phase}(a), $\Phi_{st}$ is 
nearly constant with $\Delta\widetilde{k}_{y}$ measured from the $+K$ Dirac 
point after two LZTs and $\Phi_{st}$ changes greatly with 
$\Delta\widetilde{k}_{y}$ from $-K$. Different values of the interference 
phase $\Phi_{st}$ give distinct interference patterns, such as constructive, 
destructive, and mixed quantum interference (neither constructive nor 
destructive) in the momentum space, as exemplified in 
Fig.~\ref{fig:irregular_BZO}(c).

For LZTs with a near unity transition probability, which can occur for an
infinitesimal energy gap, the $\pm K$ points make the same contribution to
the Bloch oscillations. However, slightly away from the Dirac points where
the energy gap is no longer infinitesimal, the contributions differ. The
reason why the Bloch oscillations around $K$ are more prominent compared with
those around the $-K$ point lies in the interference phase $\Phi_{st}$. In
particular, the LZTs from the $+K$ point with different momentum deviation
$\Delta\widetilde{k}_{y}$ are in phase with each other, as shown in 
Fig.~\ref{fig:stuckelberg phase}(a). However, the LZTs measured from the $-K$ 
point are out of phase, so the contributions to the Bloch oscillations cancel 
each other to some degree.

The irregular Bloch-Zener oscillations in the average current density 
integrated over the Brillouin zone arise from the LZTs induced by different 
energy gaps associated with the momentum deviation $\Delta\widetilde{k}_{y}$ 
measured from Dirac points. In the $\alpha$-$\mathcal{T}_3$ lattice, there is 
an alternative way to induce an energy gap between the valence and conduction 
bands by adding the positive (negative) onsite energy on the A (B) 
sublattice~~\cite{abranyos:2021,berman:2022}. Because of the sensitive 
dependence of the LZT on the size of the momentum-dependent energy gap, 
irregularities in the LZTs are anticipated, so are the irregular Bloch-Zener 
oscillations.

Experimentally, it may be feasible to observe at least the first peak of the 
Bloch-Zener oscillation in ballistic time~\cite{rosenstein:2010}. To probe into
the momentum-space morphology associated with LZTs and to directly observe the 
irregular Bloch-Zener oscillations in a longer time interval in Dirac material
systems remain to be difficult at the present. Alternatively, it may be 
possible to use quantum simulators~\cite{trabesinger:2012,georgescu:2014} by 
exploring equivalent optical systems~\cite{sapienza:2003}. In the past, Bloch 
oscillations have been experimentally studied in photonic 
systems~\cite{Trompeter:2006,corrielli:2013,block:2014,sun:2018,xu:2016}. 
The observation of Bloch-Zener oscillations and Landau-Zener tunneling in 
photonic graphene~\cite{sun:2018} is particularly relevant to possible 
experimental checks of our findings. In this artificial Dirac system, the wave 
packet of light is driven by an index gradient on a non-adiabatic basis and 
the two sublattices are subjected to a potential imbalance. When the momentum 
deviation is zero from a Dirac point and the index gradient is applied in a
specific direction, perfect (ideal) LZT with transition probability one can 
occur due to the zero energy gap. However, when the index gradient is applied 
in the orthogonal direction, the LZT becomes imperfect (non-ideal) due to the 
non-zero gap. Our work suggests that, if the time evolution of the wavefunction
in photonic graphene can be approximated as constituting two processes: 
adiabatic evolution and non-adiabatic LZT, in principle the St\"{u}ckelberg 
phase can be calculated to choose the appropriate index gradient value to 
create destructive interference between the quantum states in the upper and 
lower energy bands. It may thus be possible to generate destructive 
interference to obtain a near-ideal LZT. Our work predicts that, in this case, 
the resulting Bloch-Zener oscillations will become more periodic. Photonic 
graphene may be a feasible experimental testbed for these phenomena. 

\section*{Acknowledgment}

We thank Drs.~Chen-Di Han and Cheng-Zhen Wang as well as Prof.~Liang Huang for 
valuable discussions. This work was supported by AFOSR under Grant 
No.~FA9550-21-1-0186.

\appendix

\section{LZT morphology and irregular Bloch-Zener oscillations} \label{app:reason_irregular}

\begin{figure} [ht!]
\centering
\includegraphics[width=\linewidth]{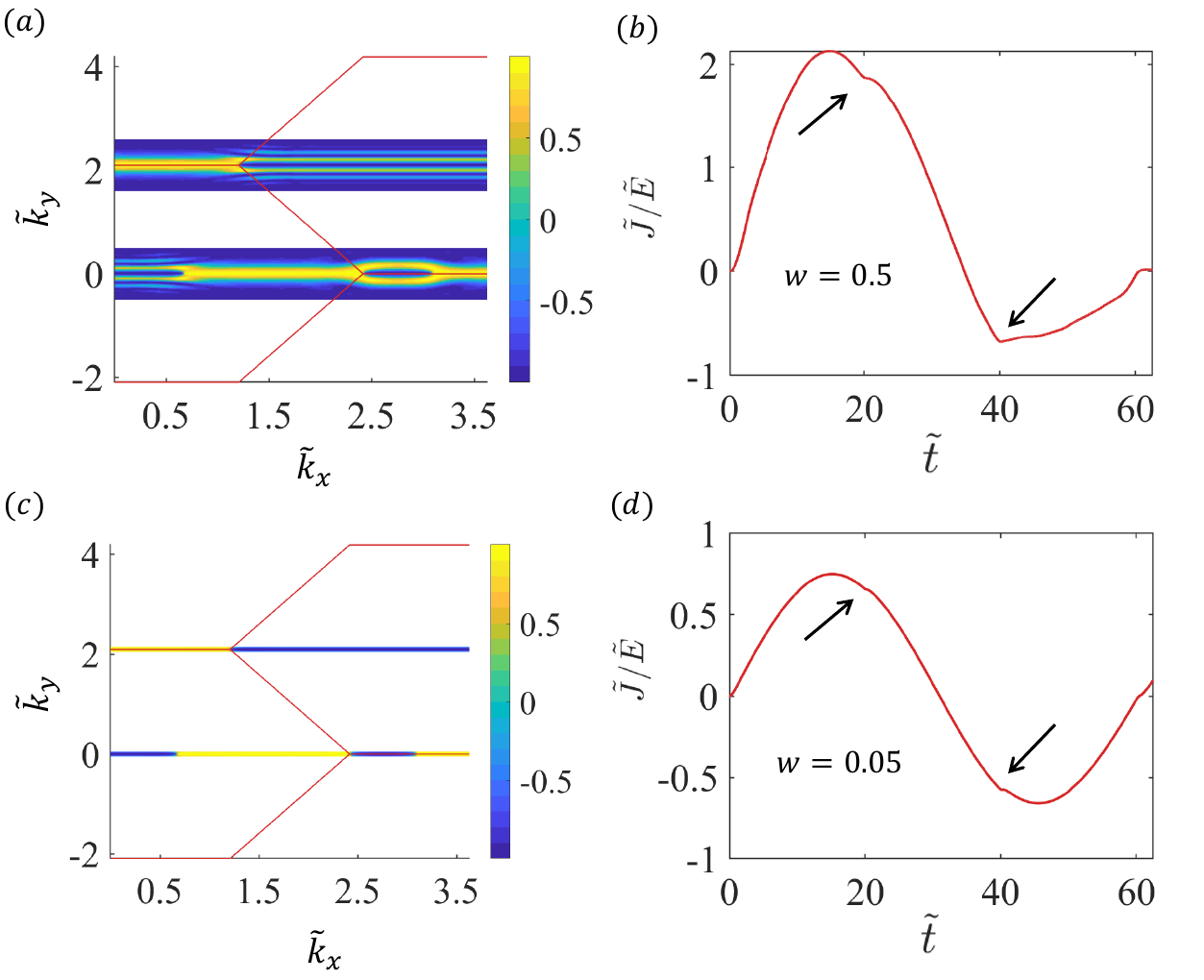}
\caption{LZTs and Bloch-Zener oscillations. (a,b) For momentum integration 
width $w=0.5$ about the $\pm K$ trace and $d\widetilde{t}\approx0.01$, the LZT 
morphology and the resulting Bloch-Zener oscillation, respectively. 
(c,d) Similar to (a,b) but for $w=0.05$ and $d\widetilde{t}\approx 0.002$. 
The irregularities in the intraband current density are indicated by arrows.
Other parameter values are 
$d\widetilde{k}_x\approx d\widetilde{k}_y\approx 0.012$ and 
$\widetilde{E}=0.1200$.}
\label{fig:reason_irregular}
\end{figure}

To further appreciate the interplay between LZT morphology and irregular 
Bloch-Zener oscillations, we consider two kinds of momentum integration 
regions about the Dirac points, as shown in 
Figs.~\ref{fig:reason_irregular}(a) and \ref{fig:reason_irregular}(c), 
respectively. The resulting time evolution of the current density is 
shown in Figs.~\ref{fig:reason_irregular}(b) and \ref{fig:reason_irregular}(d),
respectively. For the case in Fig.~\ref{fig:reason_irregular}(c), the LZTs
are near ideal, generating less irregular Bloch-Zener oscillations in 
Fig.~\ref{fig:reason_irregular}(d).

\section{Adiabatic impulse theory} \label{app:adiabatic_impulse_theory}

\subsection{Two-level systems}\label{subapp:adiabatic_impulse_two_level}

The Hamiltonian under a periodic driving in the non-adiabatic basis has the
form~\cite{shevchenko:2010}
\begin{align}
    H(t)=-\frac{\Delta}{2}\sigma_{x}-\frac{\varepsilon(t)}{2}\sigma_{z}
\end{align}
where the driving $\varepsilon(t)=\varepsilon_{0}+A\sin\omega t$ produces a 
periodic time evolution of the eigenenergy spectrum:
\begin{align}\label{eq:energy_two_level_H_LZ}
    \epsilon_{\pm}(t)=\pm\frac{1}{2}\sqrt{\Delta^2+\varepsilon(t)^2}.
\end{align}
If the bias of the driving is nonzero: $\varepsilon_0\neq0$, one period of 
the evolution of the energy contains two peaks~\cite{shevchenko:2010} that are in the time intervals
$[t_1,\;t_2]$ and $[t_2,t_1+2\pi/\omega]$, respectively. The quantum dynamical
process can be understood by using the adiabatic impulse 
theory~\cite{Damski:2006,shevchenko:2010}. According to this theory, the 
dynamical process can be approximated as the adiabatic evolution from $t_1$ 
to $t_2$ and from $t_2$ to $t_1+2\pi/\omega$, which are described by unitary 
transformation matrices $U_1$ and $U_2$, respectively, and non-adiabatic 
transitions at $t_2$ and $t_1+2\pi/\omega$ that are described by the same 
non-adiabatic transition matrix $N$. After one period, the quantum state 
under the adiabatic basis can be written as~\cite{shevchenko:2010}
\begin{align}
    \pmb{b}(t_1+2\pi/\omega)=NU_2NU_1\pmb{b}(t_1).
\end{align}
When the driving signal is linearized about the transition point as 
$\varepsilon(t)\approx -vt$, the non-adiabatic transition matrix in the 
adiabatic basis is~\cite{shevchenko:2010}
\begin{align}
    N=\left(\begin{array}{cc}
\sqrt{1-P_{LZ}}e^{-i\widetilde{\varphi}_s} & -\sqrt{P_{LZ}}\\
\sqrt{P_{LZ}} & \sqrt{1-P_{LZ}}e^{i\widetilde{\varphi}_{s}}
\end{array}\right),
\end{align}
where $P_{LZ}$ is the first LZT probability of the upper band defined by 
Eq.~\eqref{eq:formula_first_LZT_up}, $\widetilde{\varphi}_s=\varphi_s-\pi/2$, 
and $\varphi_s$ is the Stokes phase. The adiabatic evolution matrix in the 
adiabatic basis is
\begin{align}
    U_1 = \left(\begin{array}{cc}
      e^{-i\zeta_1}   & 0  \\
        0 & e^{i\zeta_{1}}
    \end{array}\right),\;\;\; U_2 = \left(\begin{array}{cc}
     e^{-i\zeta_2} & 0 \\
      0   & e^{i\zeta_2}
    \end{array}\right),
\end{align}
where $\zeta_1$ and $\zeta_2$ are the adiabatic phases given by
\begin{align}
    &\zeta_1=\int_{t_1}^{t_2}\epsilon_{+}dt,\nonumber\\
    &\zeta_2=\int_{t_2}^{t_1+2\pi/\omega}\epsilon_{+}dt.
\end{align}
In the adiabatic basis, the total transform matrix after one period is~\cite{shevchenko:2010}
\begin{align}
    NU_2NU_1=\left(\begin{array}{cc}
        \alpha & -\gamma^* \\
        \gamma & \alpha^*
    \end{array}\right),
\end{align}
where the matrix elements are 
\begin{align}
    &\gamma=2\sqrt{P_{LZ}(1-P_{LZ})}e^{-i\zeta_1}\sin(\Phi_{st}),\\
    &\alpha=e^{-i(\zeta_1+\widetilde{\varphi}_s)}[(1-2P_{LZ})\sin(\Phi_{st})+i\cos(\Phi_{st})]\nonumber,
\end{align}
and $\Phi_{st}=\zeta_2+\varphi_s$ is the St\"{u}ckelberg phase. If the initial 
state is
\begin{align}
    \pmb{b}(t_1)=(0,1)^{T},
\end{align}
the quantum state after one period will be 
\begin{align}
    \pmb{b}(t_1+2\pi/\omega)=(-\gamma^*,\;\alpha^*)^T
\end{align}
with the transition probabilities 
\begin{align}
    &P_{+} = |\gamma|^2=4P_{LZ}(1-P_{LZ})\sin^{2}(\Phi_{st}),\label{eq:two_level_p_upper}\\
    &P_{-}=|\alpha|^2=1-4P_{LZ}(1-P_{LZ})\sin^2(\Phi_{st}),
\end{align}
which satisfy the normalization constraint $P_{+}+P_{-}=1$. From 
Eq.~\eqref{eq:two_level_p_upper}, it can be seen that whether the transition 
is complete depends only on the Stokes phase and the adiabatic phase between 
two consecutive LZTs. Specifically, $\Phi_{st}=k\pi,\;k\in \mathbf{Z}$ 
corresponds to destructive interference between the quantum states in the 
upper and lower bands while $\Phi_{st}=\pi/2+k\pi$ corresponds to constructive 
interference. 

\subsection{Three-level systems}\label{subapp:adiabatic_impulse_three_level}

A periodically driven three-level system is described by
\begin{align}
    H(t)=-\frac{\Delta}{2}S_{x}-\frac{\varepsilon(t)}{2}S_{z},
\end{align}
where the time evolution of the positive and negative energy is given by
Eq.~\eqref{eq:energy_two_level_H_LZ} except for the extra flat band $\epsilon_{0}=0$. 
A previous work~\cite{carroll:1986} provided the non-adiabatic transition 
matrix in the adiabatic basis when the driving signal is linearized about the
transition point as $\varepsilon(t)\approx-vt$:
\begin{align}
    N=\left(\begin{array}{ccc}
       B+1  & A & B\\
        A & C & A\\
        B & A & B+1
    \end{array}\right),
\end{align}
where all $A$, $B$ and $C$ are constants: 
\begin{align} \nonumber
	A & \equiv-(2\widetilde{P}_{LZ})^{1/2}, \\ \nonumber
	B & \equiv P_{LZ}^{1/2}-1, \\ 
	C & \equiv 1-2P_{LZ}^{1/2},
\end{align}
and 
\begin{align} 
\widetilde{P}_{LZ}\equiv P_{LZ}^{1/2}(1-P_{LZ}^{1/2}). 
\end{align}
The adiabatic evolution matrix of the three-level system is
\begin{align}
    U_1 = \left(\begin{array}{ccc}
      e^{-i\zeta_1}   & 0  & 0\\
           0          & 1  & 0\\
           0          & 0  & e^{i\zeta_{1}}
    \end{array}\right),\; U_2 = \left(\begin{array}{ccc}
    e^{-i\zeta_2}   & 0  & 0\\
           0          & 1  & 0\\
           0          & 0  & e^{i\zeta_{2}}
    \end{array}\right).
\end{align}
The total transform matrix $M\equiv NU_2NU_1$ is
\begin{align}
    M=\left(\begin{array}{ccc}
       M_{11}    & M_{12}   & M_{13} \\
       M_{23}^*  & M_{22}   & M_{23} \\
       M_{13}^*  & M_{12}^* & M_{11}^*
    \end{array}\right),
\end{align}
where 
\begin{align}
    &M_{11}=\widetilde{(B_1+1)}^*\widetilde{(B_2+1)}^* + A\widetilde{A}^*_{1}+\widetilde{B}_{1}^*\widetilde{B}_2,\nonumber\\
    &M_{12}=A\widetilde{(B_2+1)}^*+AC+A\widetilde{B}_2,\nonumber\\
    &M_{13}=\widetilde{B}_1\widetilde{(B_2+1)}^*+A\widetilde{A}_1+\widetilde{(B_1+1)}\widetilde{B}_2,\nonumber\\
    &M_{22}=A\widetilde{A}^*_2+C^{2}+A\widetilde{A}_2,\nonumber\\
    &M_{23}=\widetilde{A}_2^*\widetilde{B}_1+\widetilde{A}_1C+\widetilde{A}_2\widetilde{(B_1+1)},
\end{align}
with 
\begin{align}
    &\widetilde{A}_i\equiv A e^{i\zeta_i},\nonumber\\
    &\widetilde{B}_{i}\equiv B e^{i\zeta_{i}},\nonumber\\
    &\widetilde{(B_{i}+1)}\equiv (B+1)e^{i\zeta_{i}},
\end{align}
for $i=1,2$. Suppose the initial quantum state is
\begin{align}
    \pmb{b}(t_1)=(0,0,1)^{T}.
\end{align}
After one period, the quantum state becomes
\begin{align}
    \pmb{b}(t_1+2\pi/\omega)=(M_{13},M_{23},M_{11}^*)^{T}.
\end{align}
The occupied probabilities of the upper, flat and lower bands after one period 
are given by
\begin{align}
    &P_{+}=16\widetilde{P}_{LZ}^2\sin^{4}(\zeta_2/2),\label{eq:three_level_upper_phase}\\
    &P_{0}=2\widetilde{P}_{LZ}((1-4\widetilde{P}_{LZ})(1-\cos\zeta_2)^2+\sin^2\zeta_2),\label{eq:three_level_flat_phase}\\
    &P_{-}=(2\widetilde{P}_{LZ}\cos\zeta_2+(1-2\widetilde{P}_{LZ}))^2,
\end{align}
respectively, where $P_{+}=|M_{13}|^2$, $P_{0}=|M_{23}|^2$ and 
$P_{-}=|M_{11}|^2$ with the constraint $P_{+}+P_{0}+P_{-}=1$. According to
Eq.~\eqref{eq:three_level_upper_phase}, for the upper band, we have that 
$\zeta_2=\pi+2k\pi,\;k\in \mathbf{Z}$ corresponds to constructive interference 
and $\zeta_2=2k\pi$ to destructive interference. For the flat band, 
$\zeta_2=2k\pi$ corresponds to destructive interference, as stipulated by
Eq.~\eqref{eq:three_level_flat_phase}. For $\zeta_2=\pi+2k\pi$, the flat band 
leads to an LZT with probability away from zero except for
$\widetilde{P}_{LZ}=1/4$ The Stokes phase defined by Eq.~\eqref{eq:stocks_phase}
depends only on the parameter $\delta$ ($0<r=2\delta<1$). Numerically, we
obtain $\varphi_s\approx 0.5$. In spite of the small phase deviation due to 
the Stokes phase $\varphi_{s}$, $\Phi_{st}=\zeta+\varphi_s=2k\pi$ corresponds
approximately to destructive interference regardless of the existence of a flat
band. However, $\Phi_{st}=\pi+2k\pi$ does not represent destructive 
interference in the three-level system. 

\section{Equivalence of $\alpha$-$\mathcal{T}_3$ lattice to strongly periodically driven two- or three-level systems} \label{app:equivalence}

When a static electric field is applied to an $\alpha$-$\mathcal{T}_3$ lattice,
the time evolution of the wavefunction can be obtained by using the adiabatic 
impulse theory~\cite{Damski:2006,shevchenko:2010}. For graphene ($\alpha = 0$)
and the dice lattice ($\alpha = 1$), the tight-binding Hamiltonians are,
respectively,
\begin{align}\label{eq:graphene_Hamiltonian}
    H=\Re\left[f_{k}(t)\right]\sigma_{x}-\Im\left[f_{k}(t)\right]\sigma_{y}
\end{align}
and 
\begin{align}\label{eq:dice_Hamiltonian}
    H=\Re\left[f_{k}(t)\right]S_{x}-\Im\left[f_{k}(t)\right]S_{y},
\end{align}
where $\sigma_x$ and $\sigma_{y}$ are the $2\times 2$ Pauli matrices for 
pseudospin-1/2 quasiparticles in graphene, $S_{x}$ and $S_{y}$ are the 
corresponding $3\times 3$ matrices from pseudospin-1 quasiparticles in the
dice lattice, and
\begin{align}
    &\Re\left[f_{k}(t)\right]=-t_\epsilon\left[1+2\cos\left(\frac{3}{2}\widetilde{k}_{y}\right)\cos\left(\frac{\text{\ensuremath{\sqrt{3}}}}{2}\widetilde{k}_{x}\left(t\right)\right)\right],\nonumber\\
    &\Im\left[f_{k}(t)\right]=2t_\epsilon\sin\left(\frac{3}{2}\widetilde{k}_{y}\right)\cos\left(\frac{\text{\ensuremath{\sqrt{3}}}}{2}\widetilde{k}_{x}\left(t\right)\right).
\end{align}
The eigenenergy spectrum of graphene is $\epsilon_{\pm}=\pm|f_{k}|$. For the 
dice lattice, there is a flat band $\epsilon_0=0$. The time evolution of 
nonzero energy band of both graphene and dice lattice has the common form 
for the upper and lower bands:
\begin{align} \label{eq:energy_equivalent}
    \widetilde{\epsilon}_{\pm}(t)=\pm\frac{1}{2}\sqrt{\Delta_{\widetilde{k}_y}^2+(\varepsilon_{0,\widetilde{k}_y}+4\sin(\omega_{\widetilde{E}}\widetilde{t}+\phi_{\widetilde{k}_x}))^2},
\end{align}
where 
\begin{align} \nonumber
\Delta_{\widetilde{k}_{y}} &=2\sin(3\widetilde{k}_y/2), \\ \nonumber
\varepsilon_{0,\widetilde{k}_y} &\equiv 2\cos(3\widetilde{k}_y/2), \\ \nonumber
\omega_{\widetilde{E}} &\equiv\sqrt{3}\widetilde{E}/2, \\ 
	\phi_{\widetilde{k}_x}&=\pi/2-\sqrt{3}\widetilde{k}_x/2.
\end{align}
Equation~\eqref{eq:energy_equivalent} has the same mathematical form as that 
for a strongly periodically driven two-level system 
[Eq.~\eqref{eq:energy_two_level_H_LZ}].

A theoretical approach to dealing with a strongly periodically driven two-level
system is the adiabatic and impulse 
approximation~\cite{Damski:2006,shevchenko:2010}, which is valid in the regime 
of strong field 
\begin{align} 
	\Delta^2+A^2\gg\omega^2 ({\rm in}\; {\rm units}\; {\rm of} \; \hbar = 1). 
\end{align}
Equations~\eqref{eq:energy_equivalent} and \eqref{eq:energy_two_level_H_LZ} 
give $A=4$ and $|\omega|=\sqrt{3}\widetilde{E}/2<0.1$, rendering applicable 
the adiabatic and impulse approximation. The idea of the analysis is to 
decompose the time evolution of system into an adiabatic evolution when it is
far from the points of avoided-crossing and non-adiabatic process in the 
vicinity of these points. 

For adiabatic evolution in graphene, the adiabatic phase in the wavefunction
depends on the integral 
$\zeta=\int \widetilde{\epsilon}_{+}\left(t\right)d\widetilde{t}$. The 
dice lattice has the same energy $\widetilde{\epsilon}_{+}$ and the flat band 
corresponds to a zero adiabatic phase $\zeta=0$. Thus, for the upper and lower 
bands in both graphene and dice lattice, the adiabatic phase has the same 
mathematical form as that for the periodically driven two- and three-level 
system, respectively.  

For the non-adiabatic transition process in graphene or dice lattice, the
effective Hamiltonian about the Dirac points is the standard 
or the three-level Landau-Zener Hamiltonian. Concretely, we can
show that the Hamiltonian for graphene and dice lattice, given by
Eqs.~\eqref{eq:graphene_Hamiltonian} and \eqref{eq:dice_Hamiltonian}, can be
written as the standard Landau-Zener Hamiltonian
for two- and three-level systems, respectively, as
\begin{align} 
H_{LZ}&=(g/2)\sigma_{x}+(st)\sigma_{z}, \\ 
H_{LZ}&=(g/2)S_{x}+(st)S_{z},
\end{align}
through some unitary transformation. In particular, the requirement is to
have $\sigma_{x}\rightarrow\sigma_{z},~\sigma_{y}\rightarrow-\sigma_{x}$ for
graphene and $S_{x}\rightarrow S_{z},~S_{y}\rightarrow- S_{x}$ for dice
lattice. These transformations can be realized by rotating the original
Hamiltonian in two steps since the physical observable does not change after
a unitary transformation. First, we rotate the Hamiltonian $H$ along the
anticlockwise direction with $\pi/2$ around the axis $y$ to get
$H\rightarrow H'$, where
$\sigma_{x}\rightarrow\sigma_{z},~\sigma_{y}\rightarrow\sigma_{y}$ for graphene
and $S_{x}\rightarrow S_{z},~S_{y}\rightarrow S_{y}$ for dice lattice. Second,
we rotate the Hamiltonian $H'$ along the anticlockwise direction with $\pi/2$
around the $z$-axis to obtain $H'\rightarrow H''$, where
$\sigma_{z}\rightarrow\sigma_{z},~\sigma_{y}\rightarrow-\sigma_{x}$ for
graphene and $S_{z}\rightarrow S_{z},~S_{y}\rightarrow-S_{x}$ for dice lattice.
The total unitary transformations for graphene and dice lattice are
\begin{align} \label{eq:graphene_rotate}
	U&=\exp\left(-i\pi/4\sigma_{y}\right)\exp\left(-i\pi/4\sigma_{z}\right),\\ \label{eq:dice_rotate} 
    U&=\exp\left(-\frac{i}{\hbar}\frac{\pi}{2}S_{y}\right)\exp\left(-\frac{i}{\hbar}\frac{\pi}{2}S_{z}\right),
\end{align}
respectively.

As for Hamiltonian expansion about Dirac points, firstly, we expand Hamiltonian in the $\widetilde{k}_{y}$ direction about the
Dirac points with 
\begin{align} \nonumber
+\mathbf{K}: \widetilde{k}_{y} & \rightarrow\delta\widetilde{k}_{y}, \\ 
-\mathbf{K}: \widetilde{k}_{y} & \rightarrow 2\pi/3+\delta\widetilde{k}_{y},
\end{align}
leading to
\begin{align} \nonumber	
\cos\left(\frac{3}{2}\widetilde{k}_{y}\right) &\approx \pm 1, \\ 
	\sin\left(\frac{3}{2}\widetilde{k}_{y}\right) &\approx\pm\frac{3}{2}\delta\widetilde{k}_{y}. 
\end{align}
With the unitary transformation in Eq.~\eqref{eq:graphene_rotate} we obtain the approximate Hamiltonian as given by
\begin{align}\label{eq:expand_H_k_y}
    U^{\dagger}\widetilde{H}U\approx-\varepsilon\left(t\right)/2\sigma_{z}-\Delta\left(t\right)/2\sigma_{x}
\end{align}
with
\begin{align}
    &\varepsilon\left(t\right)=2\left(1\pm2\cos\left(\frac{\text{\ensuremath{\sqrt{3}}}}{2}\widetilde{k}_{x}\left(t\right)\right)\right),\nonumber\\
    &\Delta\left(t\right)=\pm6\delta\widetilde{k}_{y}\cos\left(\frac{\text{\ensuremath{\sqrt{3}}}}{2}\widetilde{k}_{x}\left(t\right)\right).
\end{align}
For the dice lattice, with the total unitary transformation in Eq.~\eqref{eq:dice_rotate} we have the approximate Hamiltonian as
\begin{align}\label{eq:expand_H_k_y_three}
    U^{\dagger}\widetilde{H}U\approx-\varepsilon\left(t\right)/2 S_{z}-\Delta\left(t\right)/2 S_{x}
\end{align}
From Eqs.~\eqref{eq:expand_H_k_y} and \eqref{eq:expand_H_k_y_three}, we 
translate the momentum $\widetilde{k}_{x}$ to the Dirac points as
\begin{align} \nonumber
	+\mathbf{K}: \widetilde{k}_{x} &\rightarrow 4\pi/(3\sqrt{3})+\delta\widetilde{k}_{x}, \\ 
	-\mathbf{K}: \widetilde{k}_{x} &\rightarrow 2\pi/(3\sqrt{3})+\delta\widetilde{k}_{x}.
\end{align}
Setting $\delta\widetilde{k}_{x}\equiv\widetilde{E}\widetilde{t}^{'}$ with 
$\widetilde{t}^{'}$ starting from origin, we obtain, about $\pm\mathbf{K}$,
\begin{align}
    \cos\left(\frac{\text{\ensuremath{\sqrt{3}}}}{2}\widetilde{k}_{x}\left(t\right)\right)&\approx\left\{ \begin{array}{c}
\cos\left(\frac{2\pi}{3}-\frac{\text{\ensuremath{\sqrt{3}}}}{2}\widetilde{E}\widetilde{t}^{'}\right)\\
\cos\left(\frac{\pi}{3}-\frac{\text{\ensuremath{\sqrt{3}}}}{2}\widetilde{E}\widetilde{t}^{'}\right)
\end{array}\right\}\nonumber\\
&\approx\mp\frac{1}{2}+\frac{3}{4}\widetilde{E}\widetilde{t}^{'}.
\end{align}
where the second-order term $\delta\widetilde{k}_{x}\delta\widetilde{k}_{y}$
has been neglected. For graphene, the effective Hamiltonian about the Dirac 
points $\pm \mathbf{K}$ is
\begin{align}
    \widetilde{H}\approx\frac{3\delta\widetilde{k}_{y}}{2}\sigma_{x}\mp\frac{3\widetilde{E}\widetilde{t}'}{2}\sigma_{z},
\end{align}
which is the standard Landau-Zener Hamiltonian~\cite{wang:2017}. For the
dice lattice, the effective Hamiltonian is 
\begin{align}
    \widetilde{H}\approx\frac{3\delta\widetilde{k}_{y}}{2}S_{x}\mp\frac{3\widetilde{E}\widetilde{t}'}{2}S_{z},
\end{align}
which is the Hamiltonian of the three-level Landau-Zener 
model~\cite{carroll:1986}. In the vicinity of the Dirac points, the 
non-adiabatic Landau-Zener transition in graphene (dice lattice) induced by
a constant electric field thus shares the same quantum dynamical law as that
in the adiabatic impulse theory~\cite{Damski:2006,shevchenko:2010}. 

For the general $\alpha$-$\mathcal{T}_3$ for $\alpha \ne 0, 1$, the picture of
the quantum dynamical evolution as consisting of adiabatic evolution and 
non-adiabatic LZTs is still applicable, because the eigenenergy spectrum is 
independent of the lattice coupling parameter $\alpha$. In fact, adiabatic 
phase with $0<\alpha<1$ is the same as that for $\alpha=1$. For the
non-adiabatic process, the first LZT has been numerically calculated,
as shown in Fig.~\ref{fig:LZT_enhancement_flat}. Consequently, under the 
adiabatic impulse approximation, the dynamical evolution of the 
$\alpha$-$\mathcal{T}_3$ lattice is identical to that of the wavefunction of
a strongly periodically driven three-level system.

\section{St\"{u}ckelberg phase with electric field}\label{app:adiabatic_electric}

\begin{table} [ht!]
\caption{Electric field values at which the St\"{u}ckelberg phase is about 
$\pi$ in Fig.~\ref{fig:phaseElectric}}
\label{ta:SBZO_disruption}
\begin{tabularx}{\linewidth}{YYYYYYY}
\hline\hline
\specialrule{0em}{1pt}{1pt}
$\widetilde{t}_{B}$  &  $\widetilde{E}$ &$\widetilde{t}_{B}$  &  $\widetilde{E}$ &$\widetilde{t}_{B}$  &  $\widetilde{E}$ &  $\widetilde{E} $ weaker\\
\specialrule{0em}{1pt}{1pt}
\hline
\specialrule{0em}{1pt}{1pt}
100 &  0.0725 & 141 &  0.0514 & 193 & 0.0376 & 0.0072\\
\specialrule{0em}{1pt}{1pt}
121 &  0.06 &157 &  0.0462& 198 & 0.0366 & 0.0065\\
\specialrule{0em}{1pt}{1pt}
126 &  0.0576 &162 &  0.0448& 219 & 0.0332 & 0.006 \\
\specialrule{0em}{1pt}{1pt}
131 &  0.0554 &167 &  0.0434 & 229 & 0.0317 & 0.0059 \\
\specialrule{0em}{1pt}{1pt}
136 &  0.0533 &188 &  0.0386& 260 & 0.0279 & 0.0053 \\
\specialrule{0em}{1pt}{1pt}
\hline\hline
\end{tabularx}
\end{table}

\begin{table} [ht!]
\caption{Electric field values at which the St\"{u}ckelberg phase is about 
zero in Fig.~\ref{fig:phaseElectric}}
\label{ta:SBZO_retention}
\begin{tabularx}{\linewidth}{YYYYY}
\hline\hline
\specialrule{0em}{1pt}{1pt}
$\widetilde{t}_{B}$  &  $\widetilde{E}$ &$\widetilde{t}_{B}$  &  $\widetilde{E}$ &  $\widetilde{E}$ weaker\\
\specialrule{0em}{1pt}{1pt} \hline
\specialrule{0em}{1pt}{1pt}
118 & 0.0614 &175 & 0.0415  & 0.0069\\
\specialrule{0em}{1pt}{1pt}
144 & 0.0504 & 180 & 0.0403 & 0.0068\\
\specialrule{0em}{1pt}{1pt}
149 & 0.0487 & 185 & 0.0392 & 0.0067\\
\specialrule{0em}{1pt}{1pt}
154 & 0.0471 & 206 & 0.0352 & 0.0062\\
\specialrule{0em}{1pt}{1pt}
211 & 0.0344 & 247 & 0.0294 & 0.0042\\
\specialrule{0em}{1pt}{1pt}
216 & 0.0336 & 257 & 0.0282 & 0.0035\\
\specialrule{0em}{1pt}{1pt}
242 & 0.03 & 278 & 0.0261 & 0.003\\
\specialrule{0em}{1pt}{1pt}
\hline\hline
\end{tabularx}
\end{table}

\begin{figure} [ht!]
\centering
\includegraphics[width=\linewidth]{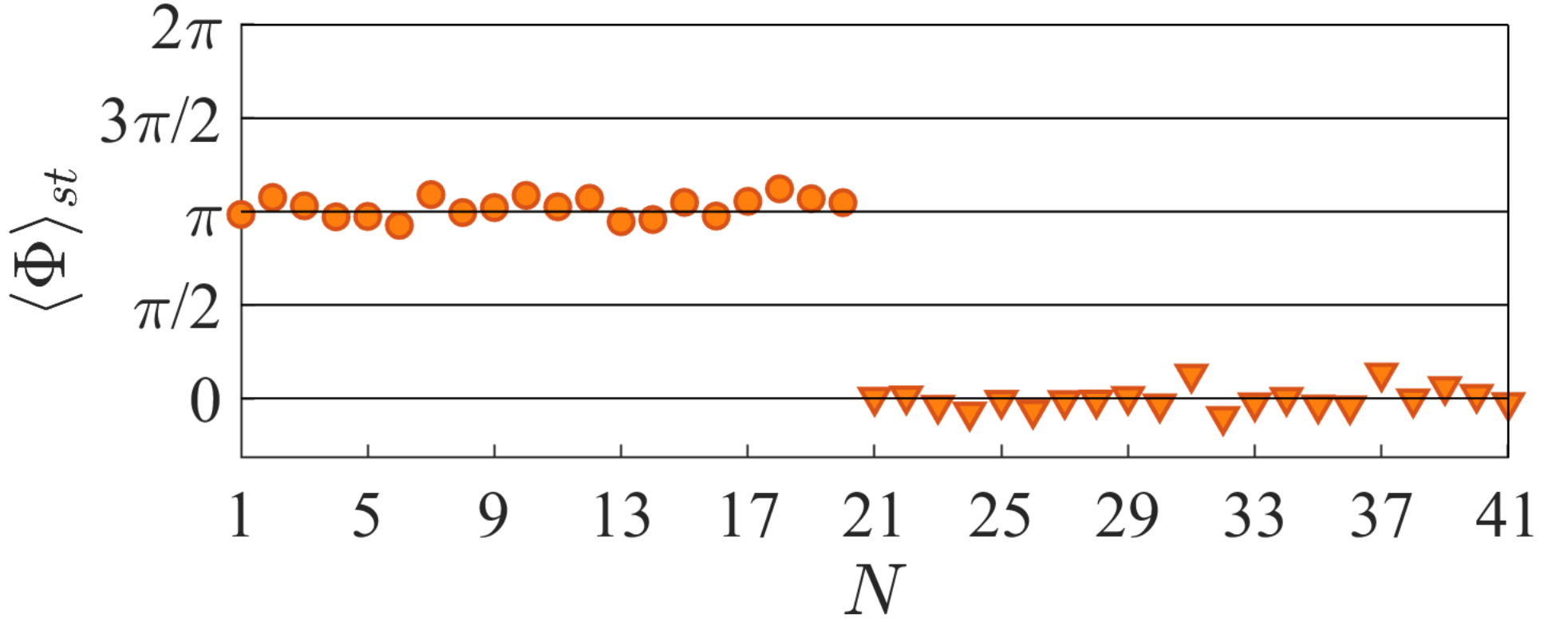}
\caption{Near zero or $\pi$ St\"{u}ckelberg phase. Shown is the 
average St\"{u}ckelberg phase over the interval of the initial momentum 
deviation $\Delta\widetilde{k}_y$ as determined by $0<r\leq1$ from the 
$+\mathbf{K}$ valley for 41 values of the electric field, where the closed 
circles are for the field values in Tab.~\ref{ta:SBZO_disruption} and the 
closed triangles correspond to the field values in Tab.~\ref{ta:SBZO_retention}.
The integer index $N$ denotes these 41 cases, with the electric field ranging 
from small to large values for both sets of data points. The integration time 
step is $d\widetilde{t}=0.01$.}
\label{fig:phaseElectric}
\end{figure}

Because the St\"{u}ckelberg phase is near constant for LZTs starting from 
the $+\mathbf{K}$ valley, as exemplified in Fig.~\ref{fig:stuckelberg phase}(a),
we define the average St\"{u}ckelberg phase over the range of the momentum 
deviation as determined by $0<r\leq1$. Although the St\"{u}ckelberg or 
adiabatic phase is sensitive to the magnitude of the electric field 
[Eqs.~(\ref{eq:adiabaticPhase}) and (\ref{eq:f_pm})], we numerically test 
two sets of electric fields to produce the average St\"{u}ckelberg phase from 
the $+\mathbf{K}$ valley around $\pi$ or zero, as shown in 
Fig.~\ref{fig:phaseElectric} for the electric field values listed in 
Tabs.~\ref{ta:SBZO_disruption} and \ref{ta:SBZO_retention} for $\alpha = 0$:
\begin{align}
\langle\zeta\rangle\approx\langle\Phi_{st}\rangle\approx\pi+2k\pi
	\ {\rm or} \ \langle\zeta\rangle \approx 2k\pi.
\end{align}
In Tabs.~\ref{ta:SBZO_disruption} and \ref{ta:SBZO_retention}, the field values
are determined based on the destructive interference pattern of
LZTs starting from the $+\mathbf{K}$ valley for $\alpha=0$ such as the orange 
traces in Figs.~\ref{fig:noflat_double_passage_like_complete}(a) 
and \ref{fig:noflat_double_passage_like_complete}(b) and are tested in the 
range of momentum deviation determined by $0<r\leq1$. For $\alpha$, two types
of behaviors can arise. In the first type, for $\alpha > 0$, for the electric 
field values from Tab.~\ref{ta:SBZO_disruption}, the LZT probability 
$|\alpha_k|^2$ can no longer reach zero after two successive LZTs, i.e., no 
destructive interference. In this case, the St\"{u}ckelberg 
phase is $\langle\Phi_{st}\rangle\approx\pi$, as shown by the filled circles
in Fig.~\ref{fig:stuckelberg phase}(b). The second type occurs for field 
values from Tab.~\ref{ta:SBZO_retention}, where $|\alpha_k|^2$ still displays 
a near-destructive interference pattern for any $\alpha > 0$. In this case, the 
St\"{u}ckelberg phase is $\langle\Phi_{st}\rangle\approx 0$, as shown by the 
filled triangles in Fig.~\ref{fig:stuckelberg phase}(b). 

\bibliography{LZT_BO}

\end{document}